\documentclass[usenatbib]{mn2e}

\usepackage[dvips]{graphicx}
\usepackage{amssymb}
\usepackage{txfonts}
\usepackage{colordvi}

\newcommand{\xte}{{\textit{RXTE}}}

\newcommand{\sax}{{\textit{Beppo\-SAX}}}
\newcommand{\gro}{{\textit{CGRO}}}

\newcommand{\swift}{{\textit{Swift}}}

\newbox\grsign \setbox\grsign=\hbox{$>$} \newdimen\grdimen \grdimen=\ht\grsign
\newbox\simlessbox \newbox\simgreatbox \newbox\simpropbox
\setbox\simgreatbox=\hbox{\raise.5ex\hbox{$>$}\llap
     {\lower.5ex\hbox{$\sim$}}}\ht1=\grdimen\dp1=0pt
\setbox\simlessbox=\hbox{\raise.5ex\hbox{$<$}\llap
     {\lower.5ex\hbox{$\sim$}}}\ht2=\grdimen\dp2=0pt
\setbox\simpropbox=\hbox{\raise.5ex\hbox{$\propto$}\llap
     {\lower.5ex\hbox{$\sim$}}}\ht2=\grdimen\dp2=0pt
\def\ga{\mathrel{\copy\simgreatbox}}

\def\simprop{\mathrel{\copy\simpropbox}}

\title[Radio/X-ray correlations in Cyg X-1]{X-ray variability patterns and radio/X-ray correlations in Cyg X-1}

\author[A. A. Zdziarski et al.]
{Andrzej A. Zdziarski,$^1$\thanks{E-mail: aaz@camk.edu.pl} Gerald K. Skinner,$^{2,3}$ Guy G. Pooley,$^4$ and Piotr Lubi\'nski$^5$\\
$^1$Centrum Astronomiczne im.\ M. Kopernika, Bartycka 18, PL-00-716 Warszawa, Poland\\
$^2$Astroparticle Physics Laboratory, Code 661, CRESST and NASA/Goddard Space Flight Center, Greenbelt, MD 20771, USA\\ 
$^3$Department of Astronomy, University of Maryland, College Park, MD 20742, USA\\
$^4$Cavendish Laboratory, J. J. Thomson Avenue, Cambridge CB3 0HE\\
$^5$Centrum Astronomiczne im.\ M. Kopernika, Rabia\'nska 8, PL-87-100 Toru\'n, Poland\\
}

\date{Accepted 2011 May 24.  Received 2011 May 22; in original form 2011 March 15}

\pagerange{\pageref{firstpage}--\pageref{lastpage}}
\pubyear{2011}

\begin{document}

\maketitle

\label{firstpage}

\begin{abstract}
We have studied the X-ray variability patterns and correlations of the radio and X-ray fluxes in all spectral states of Cyg X-1 using X-ray data from \xte/ASM, \gro/BATSE, and \swift/BAT. In the hard state, the dominant spectral variability is a changing of normalisation with fixed spectral shape, while in the intermediate state the slope changes, with a pivot point around 10 keV. In the soft state, the low energy X-ray emission dominates the bolometric flux which is only loosely correlated with the high energy emission. In black hole binaries in the hard state, the radio flux is generally found to depend on a power of the X-ray flux, $F_{\rm R}\propto F_{\rm X}^p$. We confirm this  for Cyg X-1. Our new finding is that this correlation extends to the intermediate and soft states provided the broad-band X-ray flux in the Comptonization part of the spectrum (excluding the blackbody component) is considered instead of a narrow-band medium-energy X-ray flux. We find an index $p\simeq 1.7\pm 0.1$ for 15 GHz radio emission, decreasing to $p\simeq 1.5\pm 0.1$ at 2.25 GHz. We conclude that the higher value at 15 GHz is due to the effect of free-free absorption in the wind from the companion. The intrinsic correlation index remains uncertain. However, based on a theoretical model of the wind in Cyg X-1, it appears to be close to $\simeq$1.3, which, in the framework of accretion/jet models, implies that the accretion flow in Cyg X-1 is radiatively efficient. The correlation with the flux due to Comptonization emission indicates that the radio jet is launched by the hot electrons in the accretion flow in all spectral states of Cyg X-1. On the other hand, we rule out the X-ray jet model. Finally, we find that the index of the correlation, when measured using the X-ray flux in a narrow energy band, strongly depends on the band chosen and is, in general, different from that for either the bolometric flux or the flux in the hot-electron emission. 
\end{abstract}

\section{Introduction}
\label{intro}

The relationship between the radio and X-ray emission in black-hole binaries has been extensively studied, see, e.g., \citet*{gfp03}, \citet*{mhd03}, \citet{corbel00,corbel03,corbel04,corbel08}, \citet{yc05}, \citet{soleri10}, \citet{coriat11}. The qualitative form of the radio/X-ray correlation in Cyg X-1 is rather typical of black-hole binaries. In the hard spectral state, there is a positive correlation, as in other black-hole binaries \citep{corbel00,corbel04,gfp03}. At transitions to the soft state of Cyg X-1, the radio flux strongly decreases. This is also similar to other black-hole binaries. In particular, the radio flux strongly declines at the transition from the hard state to a soft one in GX 339--4 \citep{corbel00}, XTE J1650--500 \citep{corbel04}, and XTE J1859+226 \citep*{fbg04}. After the soft-state decline, the radio flux from black-hole transients may strongly increase in an intermittent manner, sometimes up to the overall maximum for a source, e.g., \citet{fbg04}, \citet{gallo04}, which is probably due to the ejection of radio-emitting blobs (e.g., \citealt{y09}). A similar overall radio-soft X-ray correlation with all of the above states is also seen in Cyg X-3 \citep*{szm08}, which, based on its X-ray properties \citep{sz08,hj09}, is probably a black-hole binary. The soft-state radio flaring is not seen in Cyg X-1, apparently due to the limited range of its X-ray flux. 

Determining the relationship between the radio and X-ray emission is of major importance to our understanding of the physics of accretion and outflow in black-hole binaries. The bulk of the radio emission is clearly from the jet. There are two possible origins of the correlation of the radio emission with X-rays in black-hole binaries. One is that the level of X-ray emission in the hard spectral state, related to the accretion rate, is in turn related to the rate of the outflow forming the jet (e.g., \citealt*{mirabel98,corbel00,hs03,ycn05}). Another is that the X-ray emission of black-hole binaries is dominated by emission of the jet (e.g., \citealt*{vadawale01,georganopoulos02,markoff03,falcke04}). 

The usual approach to studying the radio/X-ray correlation is to consider the hard spectral state only and to study the X-ray emission in a relatively narrow band, typically 2--10 keV or 3--9 keV. The relation with the radio flux is represented by a power law (e.g., \citealt{gfp03,mhd03,coriat11}) and theoretical consequences of the value of the power-law index are then are considered (e.g., \citealt{mhd03,hs03,coriat11}). Sometimes infrared or optical emission is also considered in addition to radio (e.g., \citealt{coriat09}). A common implicit assumption is that the correlation is the same in other X-ray bands as in the band considered. Attention is rarely paid to the correlation of the radio flux with the bolometric flux (dominated by the X-rays), a quantity which is crucial for the first class of models listed above. Here, we consider the correlation of the radio emission with X-rays in several energy bands in the case of the archetypical and well studied black-hole binary Cyg X-1. We consider all its spectral states, from hard to soft. 

We first study the X-ray spectral variability patterns, in particular the dependence of the bolometric flux on the X-ray spectral slope, as well as on the fluxes in various X-ray bands. We calculate the spectra and bolometric fluxes averaged over a 1-day time scale using the monitoring by the All-Sky Monitor (ASM) on board {\it Rossi X-ray Timing Explorer\/} (\xte; \citealt*{brs93,levine96}) simultaneous with the monitoring either by the \swift\/ Burst Alert Telescope (BAT; \citealt{barthelmy05,m05}) or by the Burst and Transient Source Experiment (BATSE; \citealt{harmon02}) on board of {\it Compton Gamma Ray Observatory\/} (\gro). 

The discovery of substantial orbital modulation of the radio emission \citep*{pfb99} shows that free-free absorption in the stellar wind \citep{gies03} of the donor affects the observed radio emission and so complicates the interpretation of the observations. The level of attenuation of the intrinsic radio flux depends on the height along the jet where the bulk of the emission at a given frequency emerges \citep{sz07}, which is difficult to determine. In spite of this complication, given its importance and brightness and the wealth of available monitoring data, the study of Cyg X-1 source provides valuable insights into the relationship between the radio and X-ray emission in black-hole binaries.

\begin{figure}
\centerline{\includegraphics[height=\columnwidth,angle=-90]{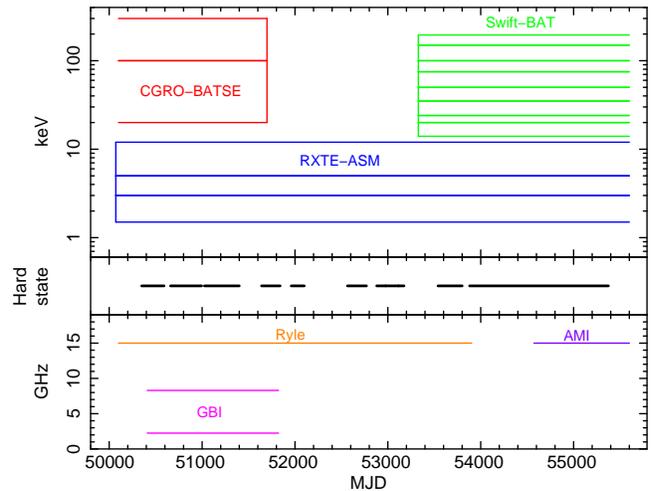}}
\caption{The time and photon energy/wavelength coverage by the instruments used in this work. The periods dominated by the hard state are indicated by the horizontal black lines.
}
\label{f:coverage}
\end{figure}

\section{The data}
\label{data}

We use monitoring data from the \xte/ASM, which has three channels at energies of 1.5--3 keV, 3--5 keV and 5--12 keV. The \gro/BATSE monitoring data used here are the same as those in \citet{z02}, hereafter Z02, which were obtained using the Earth occultation analysis technique \citep{harmon02}. They are given as energy fluxes in the 20--100 keV and 100--300 keV channels. For the \swift/BAT, we use 14--195 keV 8-channel light curves created for this work. The channels are between energies of 14, 20, 24, 35, 50, 75, 100, 150 and 195 keV. The data used here from the ASM, BATSE and BAT are for MJD 50087--55700, 50087--51686 and 53355--55469, respectively. Fig.\ \ref{f:coverage} shows the time and energy coverage of the instruments used here.

The ASM count rates are converted into energy fluxes using the matrix of Z02. We convert the BAT 8-channel data into energy fluxes by scaling to the Crab spectrum using the results of \citet{rv01} up to 100 keV, and at higher energies, those of \citet{jr09}, see \citet*{z11}, hereafter ZPS11, for details.

\begin{figure*}
\centerline{\includegraphics[width=15cm]{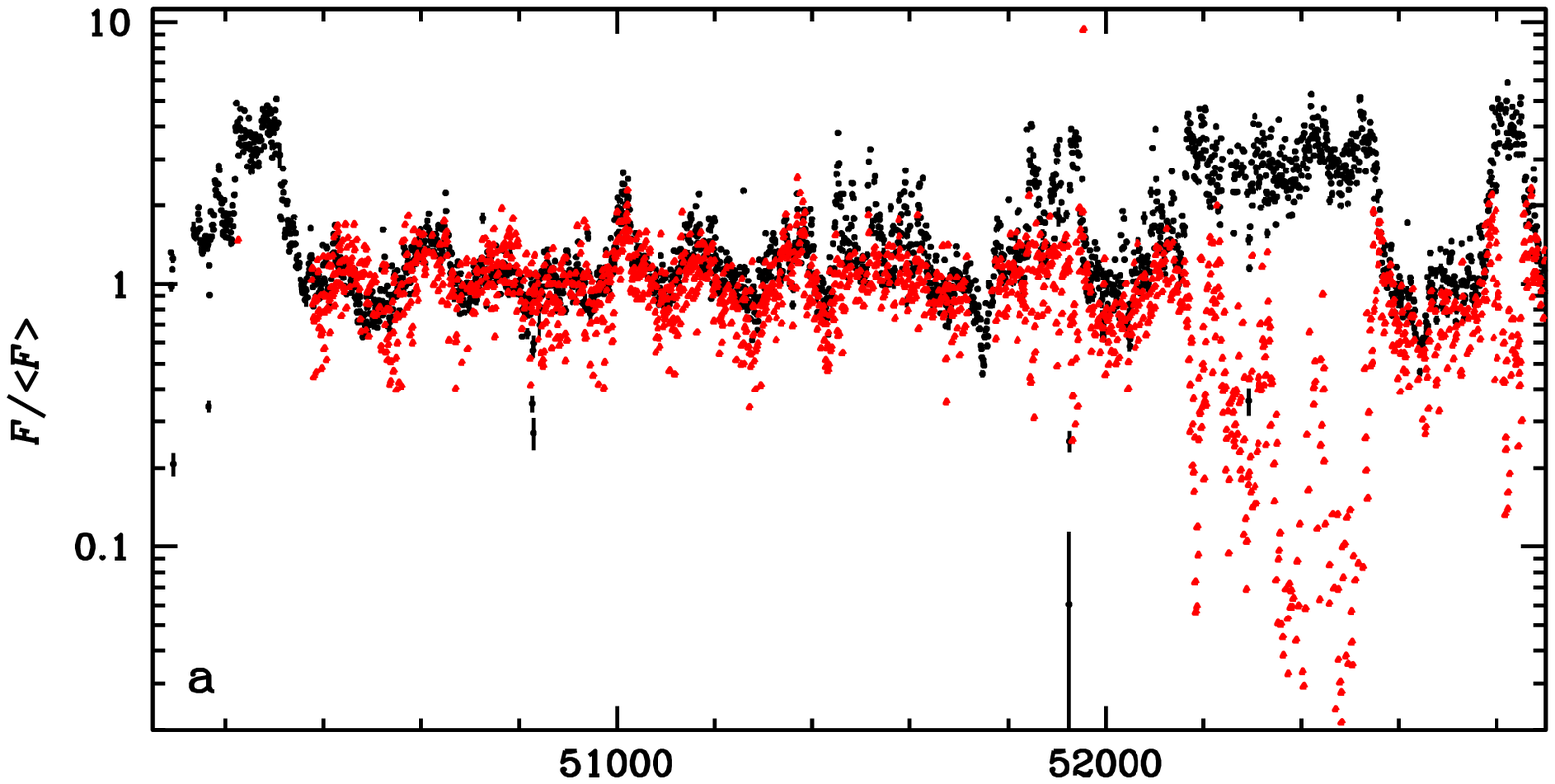}} 
\centerline{\includegraphics[width=15cm]{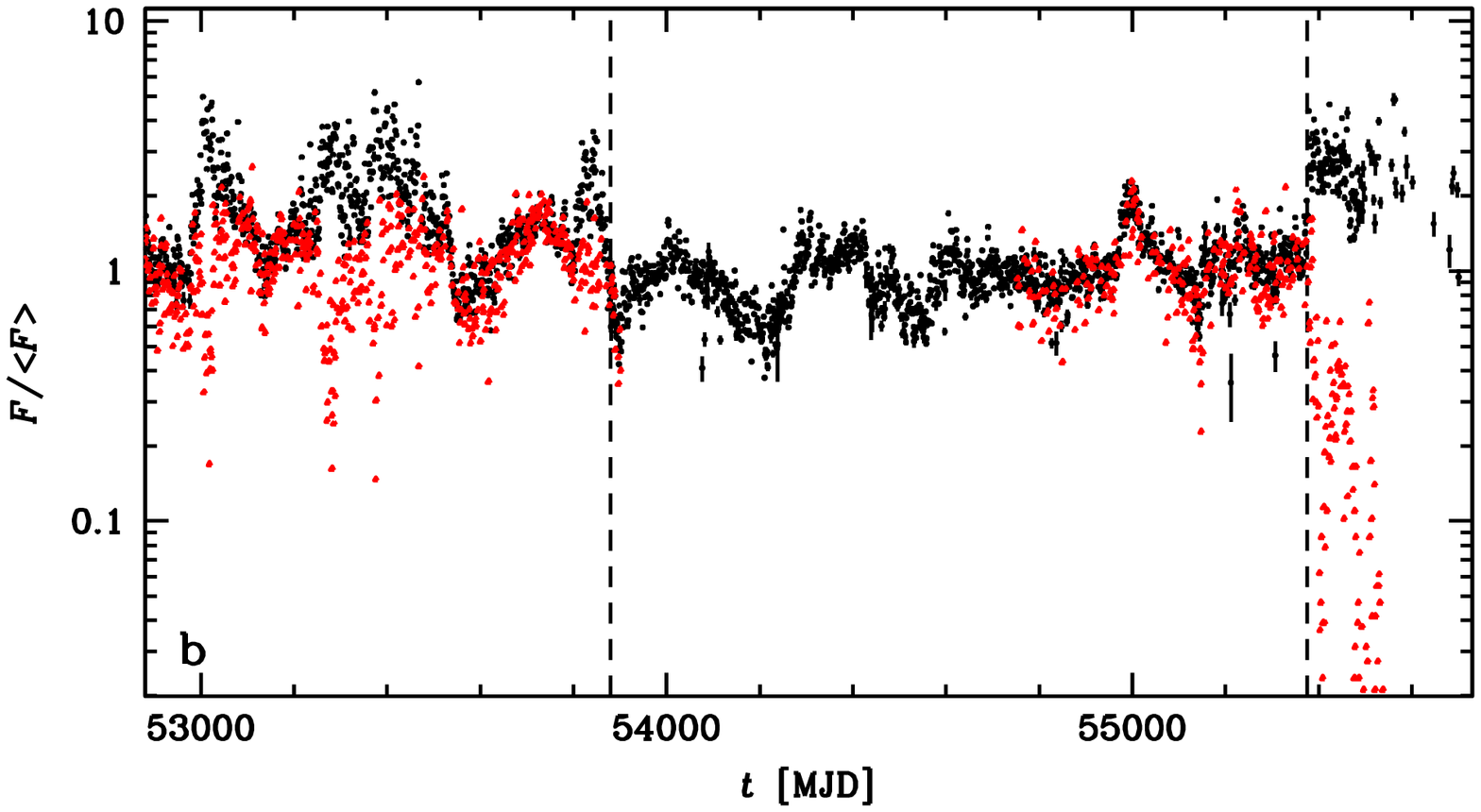}} 
\caption{The light curves showing daily averages from the ASM (black error bars, 1.5--12 keV) and Ryle/AMI (red points, 15 GHz), normalized to their respective (un-weighted) average values within the hard state of MJD 53880--55375, an interval delineated in panel (b) by the vertical dashed lines. The averages are $\langle F\rangle= 20.7$ s$^{-1}$ and 12.7 mJy, respectively. For clarity, the uncertainty of the radio points is not shown. During some soft states, the radio fluxes were below the range of $F/\langle F\rangle$ shown.  The 2010/11 soft state can be seen in  panel (b).
}
\label{f:lc}
\end{figure*}

\begin{figure}
\centerline{\includegraphics[width=7cm]{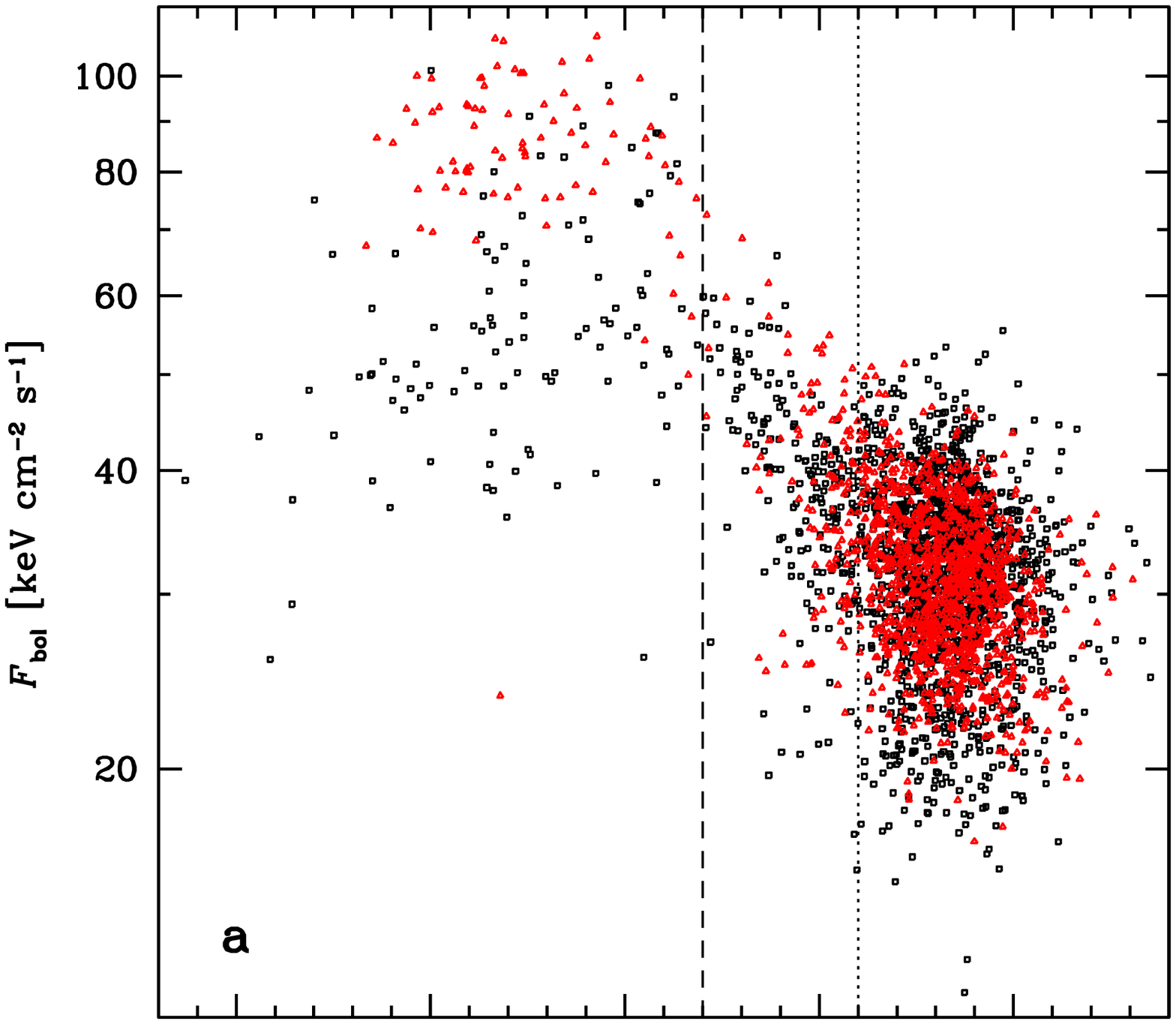}} 
\centerline{\includegraphics[width=7cm]{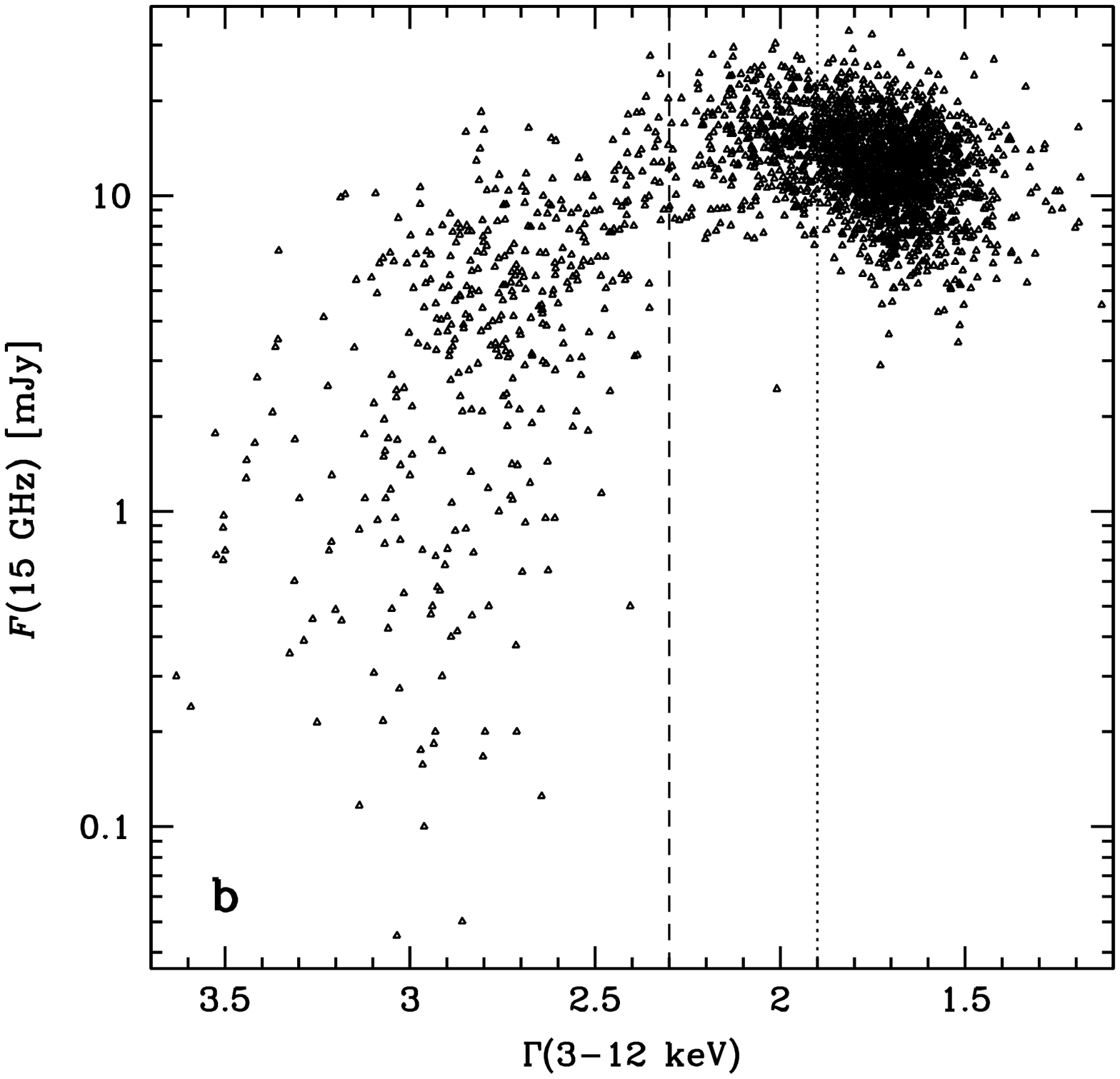}} 
\caption{Relationship between the 3--12 keV photon index calculated from the ASM data and (a) the bolometric flux calculated from the ASM and BATSE data (red points) and from the ASM and BAT data (black points), and (b) the daily-averaged radio flux at 15 GHz. The dotted and dashed vertical lines show our assumed boundaries between the hard and intermediate states, and intermediate and soft states, respectively. For clarity of display, we do not plot the error bars and include only the data with low index uncertainty, $\Delta \Gamma(3$--$12\, {\rm keV})<0.3$. The decreasing value of $\Gamma$ on the horizontal axis corresponds to the increasing spectral hardness.
}
\label{f:gamma}
\end{figure}

We divide the daily data from the ASM and those from other instruments simultaneous with the ASM monitoring into three groups based on their 3--12 keV photon index, $\Gamma$, which we calculate using the method of Z02 and ZPS11. The 3--12 keV energy range in Cyg X-1 is only weakly affected by absorption and is thus a good measure of the actual spectral slope in that range. We consider separately data with $\Gamma<1.9$ (hard state), $1.9<\Gamma<2.3$ (intermediate state) and $\Gamma>2.3$ (soft state). These boundaries differ slightly from those used in Z02; here, in particular, our definition of the hard state is more stringent and more suitable for studies of the correlation with the radio emission. We take $\Gamma_{\rm av}=1.7$, 2.1 and 2.5 as the average indices of the three respective groups. We stress that the soft state in Cyg X-1 usually contains a high energy tail much stronger than that typical of low-mass black-hole binaries in the soft state. In those binaries, the tail often virtually disappears (e.g., \citealt{gd04}) as well as the radio emission being strongly quenched (unlike the case in Cyg X-1). Thus the soft state of Cyg X-1 might also be called `soft intermediate' in the context of low-mass black-hole binaries. 

We use 15-GHz data from the Ryle Telescope and the AMI Large Array. The AMI Large Array is the re-built and reconfigured Ryle Telescope. The monitoring by the Ryle Telescope was carried out during MJD 50226--53902, and the AMI monitoring of Cyg X-1 started on MJD 54573,  the data used here being until MJD 55540. The new correlator has a useful bandwidth of about 4 GHz (compared with 0.35 GHz); but the effective centre frequency is similar, and in any case the radio spectrum of Cyg X-1, at least in the hard state, is known to be very flat \citep{fender00}. The data are subject to variations in the flux calibration of about 10 per cent from one day to another.

We also use the 2.25 GHz and 8.3 GHz monitoring data\footnote{ftp://ftp.gb.nrao.edu/pub/fghigo/gbidata/gdata/} from the Green Bank Interferometer (GBI). According to the web page description, the data are error-dominated below 15 mJy. However, we have found that using only the data $>F_{\rm R,min}=15$ mJy with the X-ray fluxes introduces a bias, making the resulting radio/X-ray distributions artificially flat. In fact, the values of the Spearman's correlation coefficients increase and the associated null hypothesis probabilities decrease when we decrease $F_{\rm R,min}$ (i.e., include more low-flux data). After studying different choices, we have opted for using the data $>F_{\rm R,min}=10$ mJy for fits. However, we show all the radio points in the figures. The GBI monitoring of Cyg X-1 lasted MJD 50409--51823.

\section{Results}
\label{results}

\subsection{The light curves}
\label{lc}

Fig.\ \ref{f:lc} presents the daily-average light curves from the ASM (1.5--12 keV) and Ryle/AMI (15 GHz). Both have been renormalized to the average count rate/flux over the long hard state of MJD 53880--55375 (discussed in detail by ZPS11). We see that the radio and the 1.5--12 keV X-ray fluxes closely follow each other during hard/low X-ray states but the radio emission is suppressed during the soft/high X-ray states. The radio emission during the latter states can become extremely low, but also can flare during the soft state back to the hard-state level, e.g., during the long soft state of MJD 52160--52570. Fig.\ 1 in ZPS11 shows also the 15--50 keV BAT light curve, normalized in the same way. We quantify the radio/X-ray correlation in Section \ref{corr} below.

During the 15 GHz monitoring of Cyg X-1, there were four events when the flux rose above 50 mJy. We show their light curves in detail in Appendix \ref{flares}. 

\subsection{X-ray spectra and variability patterns}
\label{xray}

Appendix \ref{bol} gives details of our method of calculating the bolometric flux (i.e., that integrated over all energies) based on the ASM, BATSE and BAT data. Fig.\ \ref{f:gamma}(a) shows the relationship between the resulting bolometric flux and $\Gamma(3$--12 keV). In order to allow comparison with commonly plotted hardness-flux diagrams, the horizontal axis shows $\Gamma$ decreasing to the right. We see a positive correlation of $F_{\rm bol}$ with $\Gamma$ in the intermediate state, with the Spearman's rank-order correlation coefficient, $r_{\rm s}\simeq 0.16$, and the corresponding probability that this $r_{\rm s}$ is due to a chance, $P_{\rm s}\simeq 1\times 10^{-14}$. The correlation continues to the hard state for $F_{\rm bol}>30$ keV cm$^{-2}$ s$^{-1}$, within which $r_{\rm s}\simeq 0.52$, $P_{\rm s}\simeq 2\times 10^{-17}$. In the hard state at lower $F_{\rm bol}$, it changes by a factor of a few at a given $\Gamma$ without any apparent trend. 

\begin{figure*}
\centerline{\includegraphics[height=6.3cm]{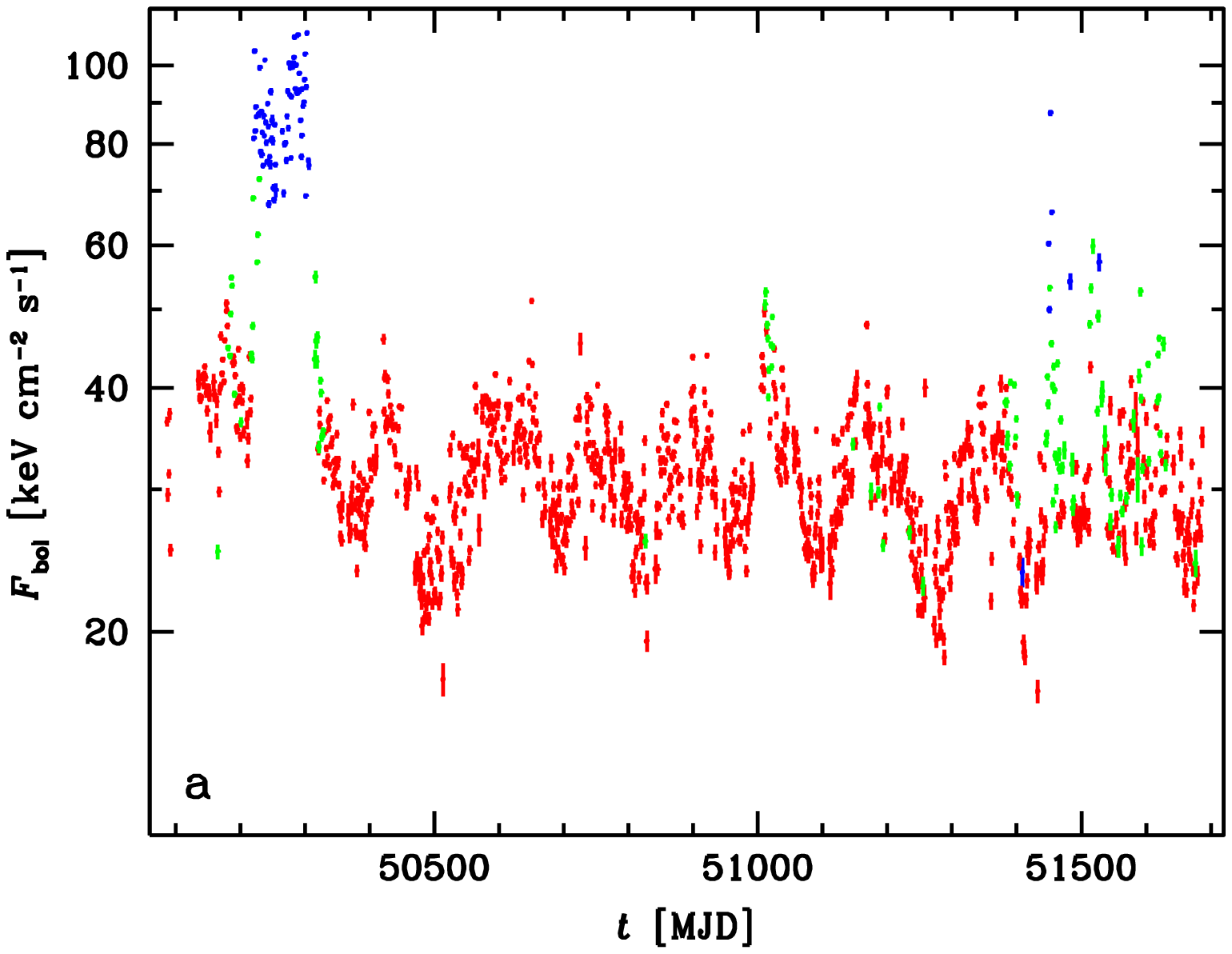} 
\includegraphics[height=6.3cm]{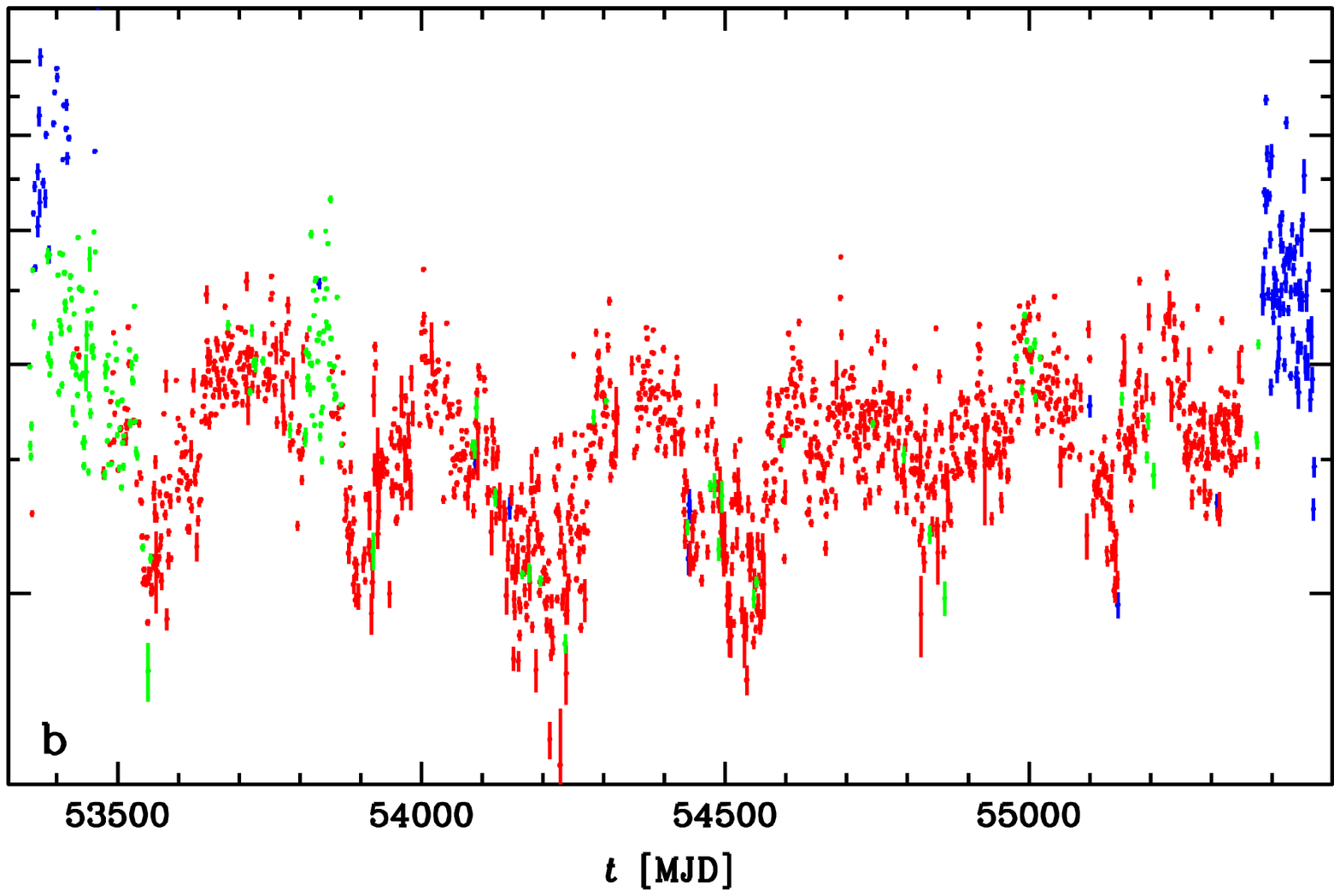}} 
\caption{The light curves showing daily averages of the bolometric flux, estimated based on (a) the ASM+BATSE data, and (b) the ASM+BAT data. The red, green and blue symbols correspond to $\Gamma(3$--$12\,{\rm keV}) <1.9$ (hard state), $1.9<\Gamma<2.3$ (intermediate state) and $\Gamma>2.3$ (soft state), respectively.
}
\label{f:lc_bol}
\end{figure*}

The soft state data, at $\Gamma(3$--12 keV$)\ga 2.3$, show no clear overall correlation. In this state, $\Gamma(3$--12 keV) is not uniquely correlated with the bolometric flux. We note that in a number of occurrences of the soft state $F_{\rm bol}$ was lower than that in the high-flux hard state. This happened mostly during the 2010/11 soft state, when the ASM fluxes reached relatively low values, see Figs.\ \ref{f:lc}(b), \ref{f:lc_bol}(b).

Fig.\ \ref{f:lc_bol} presents the light curve of $F_{\rm bol}$. We see changes up to a factor of $\sim$10. We  clearly see the superorbital periodicities of Cyg X-1, $\sim$150 d in Fig.\ \ref{f:lc_bol}(a) and $\sim$300 d in Fig.\ \ref{f:lc_bol}(b), see, e.g., \citet*{pzi08}, ZPS11.

\begin{figure*}
\centerline{
\includegraphics[height=3.5cm]{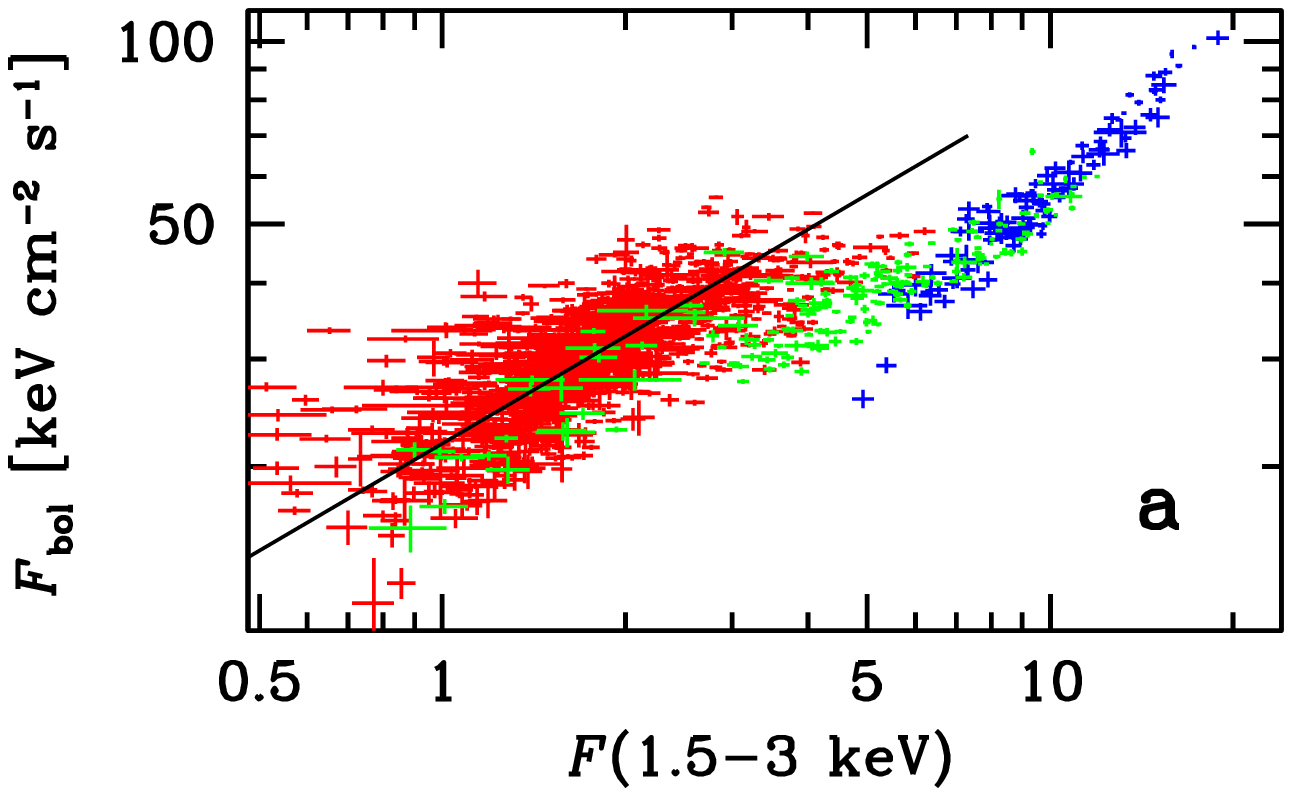}
\includegraphics[height=3.5cm]{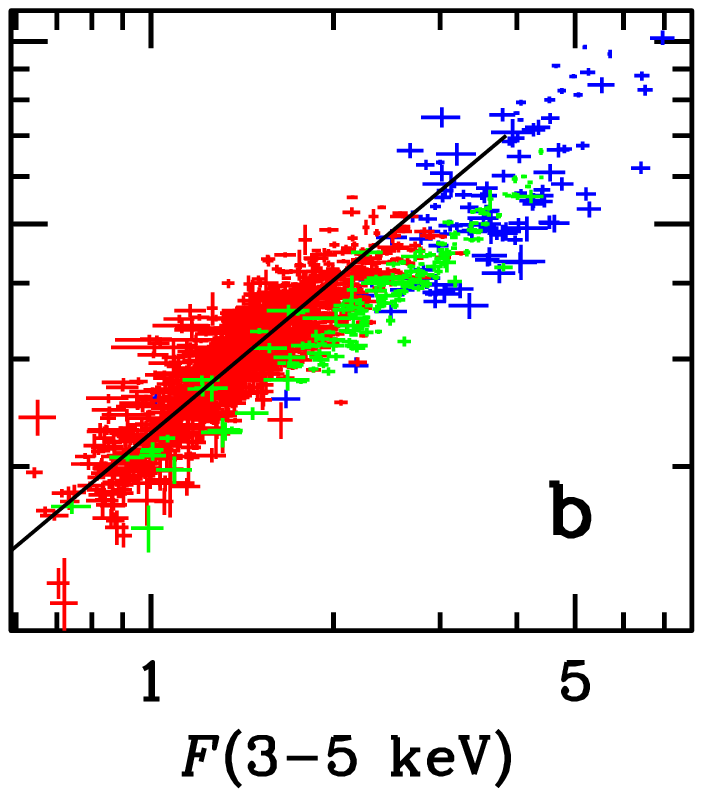}
\includegraphics[height=3.5cm]{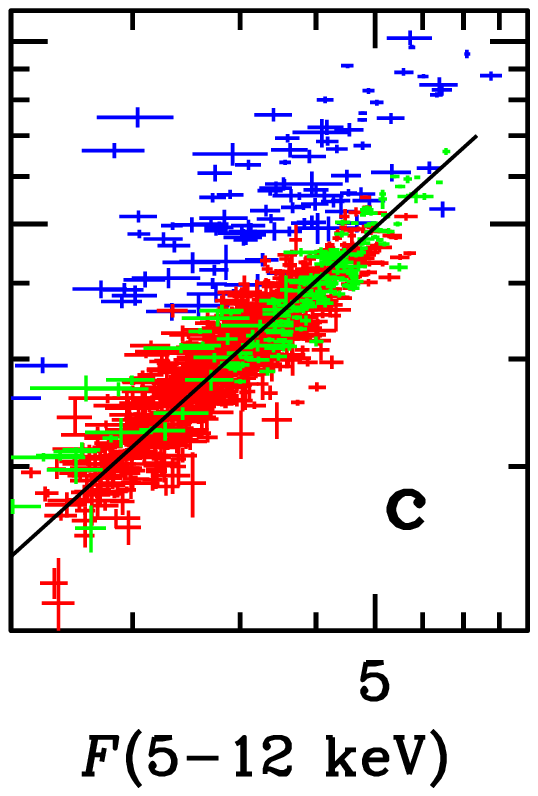}
\includegraphics[height=3.5cm]{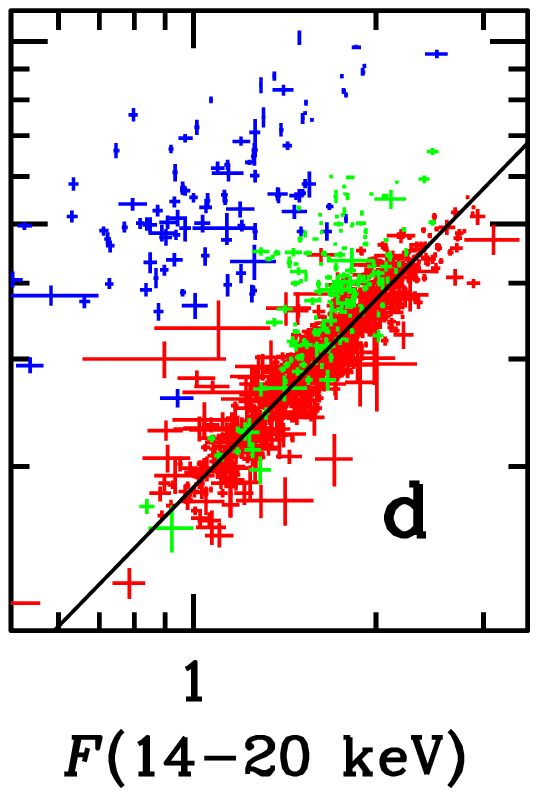}
\includegraphics[height=3.5cm]{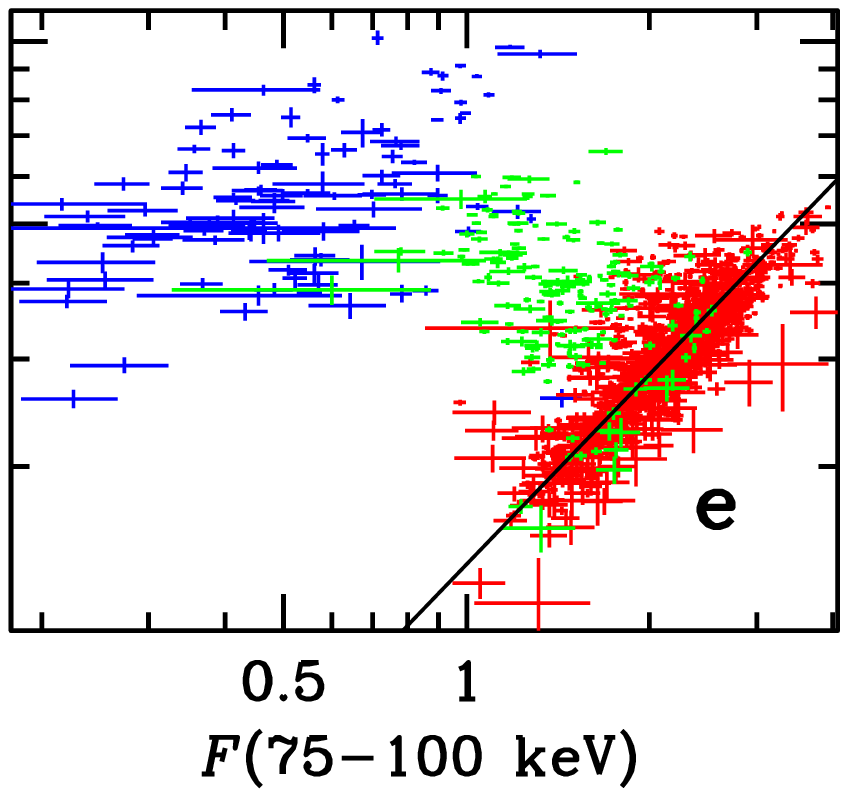}}
\centerline{
\includegraphics[height=7.81cm]{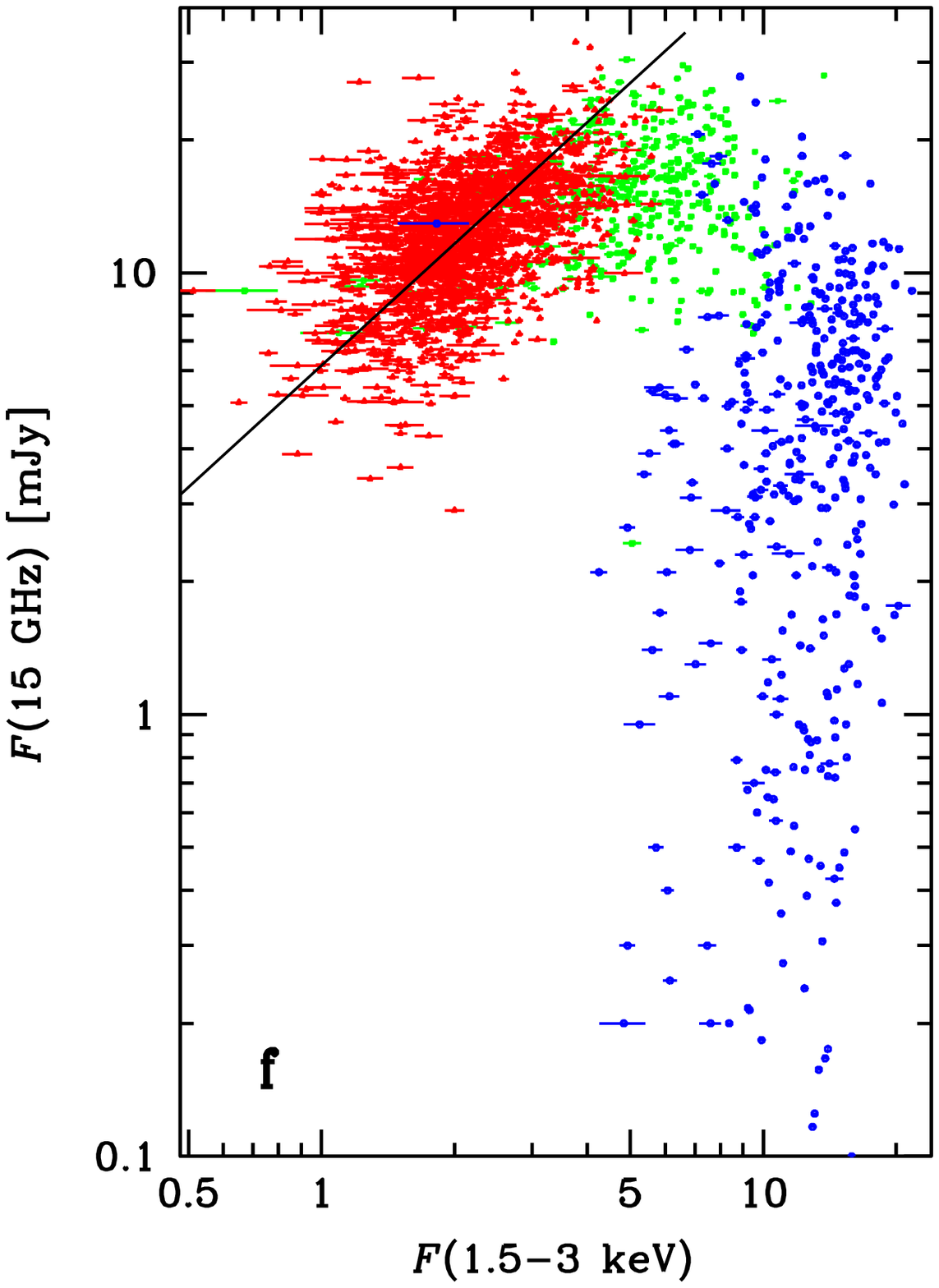}
\includegraphics[height=7.81cm]{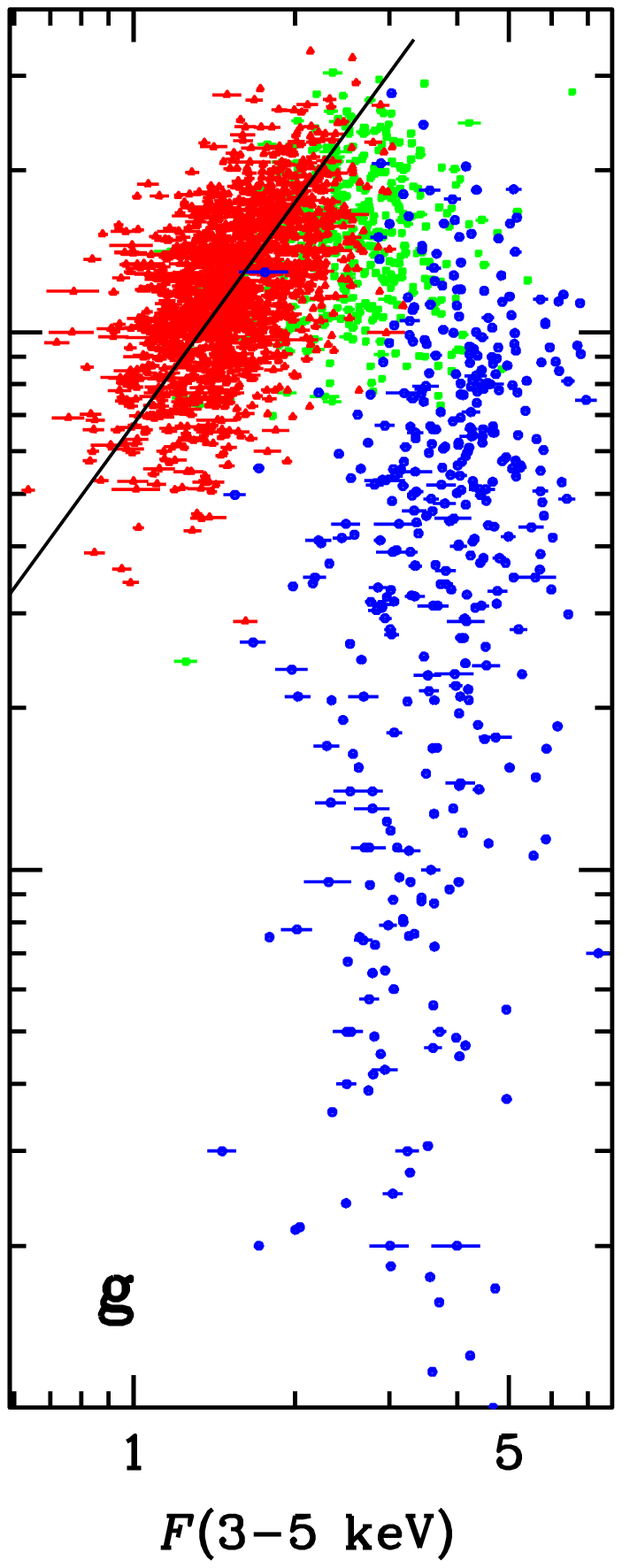}
\includegraphics[height=7.81cm]{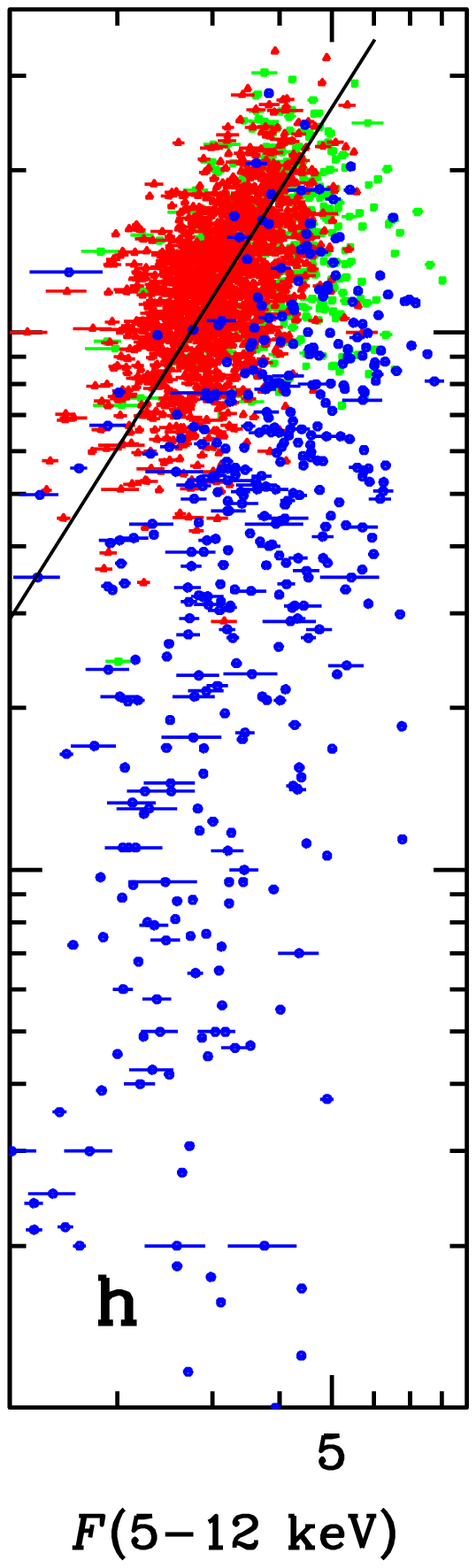}
\includegraphics[height=7.81cm]{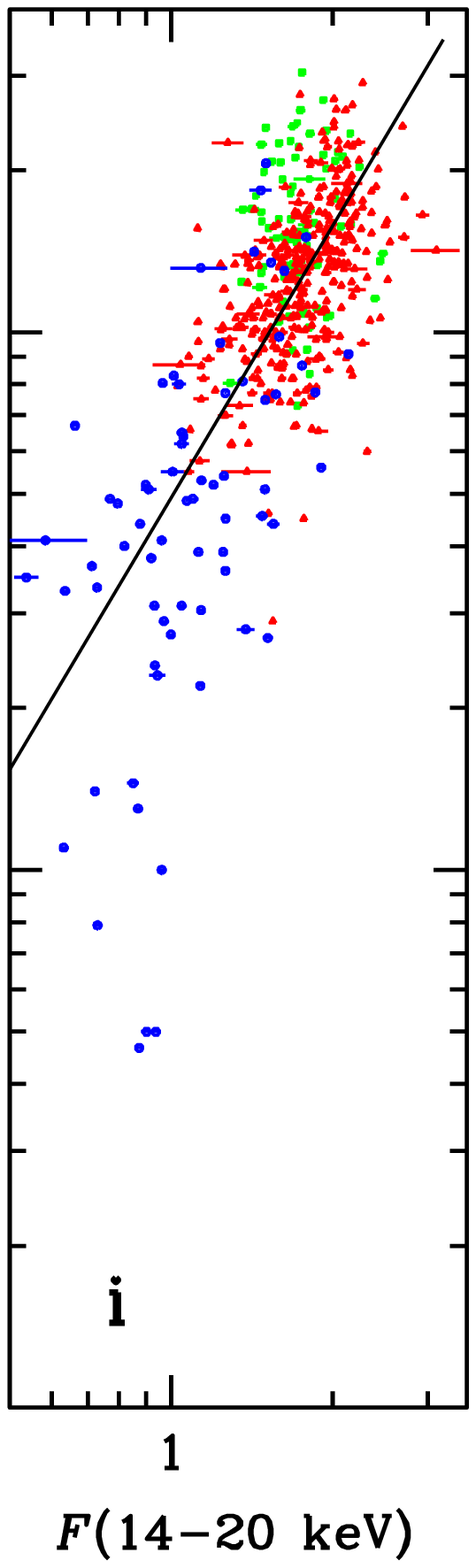}
\includegraphics[height=7.81cm]{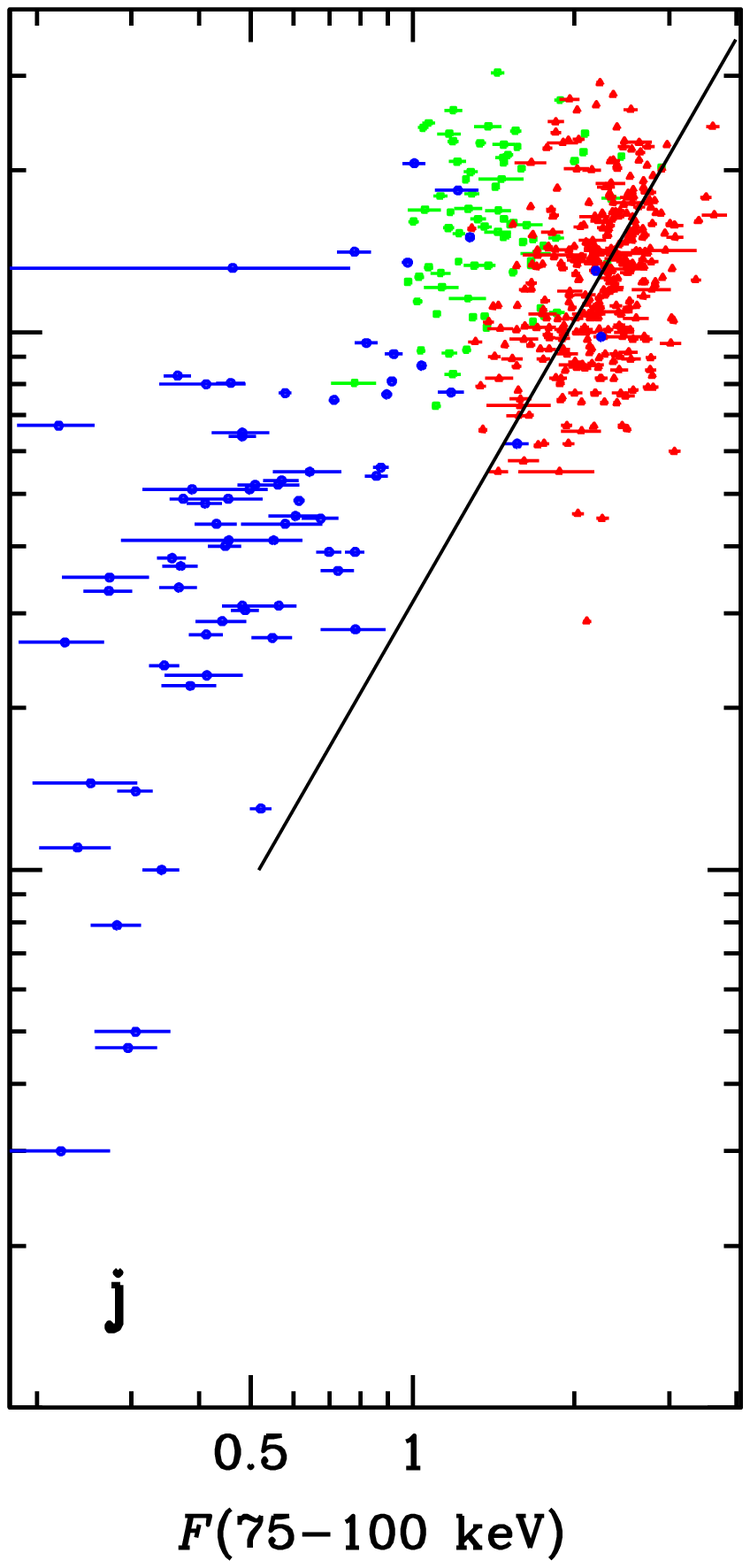}}
\caption{Top panels (a--e): the X-ray variability patterns of Cyg X-1 as illustrated by the relationship between the bolometric flux (calculated using the ASM+BAT data) and the flux in a given band. Bottom panels (f--j): the relationship between the 15 GHz radio flux and the X-ray fluxes measured by either the ASM ($\leq 12$ keV) or BAT ($\geq 14$ keV) (daily averages). 
The red, green and blue symbols correspond to data for which the 3--12 keV photon index is $\Gamma<1.9$ (hard state), $1.9<\Gamma<2.3$ (intermediate state) and $\Gamma>2.3$ (soft state), respectively. The lines show the best-fit power-law fits [equation (\ref{fit})] to the hard state data. The units of the X-ray flux are keV cm$^{-2}$ s$^{-1}$ and the error bars on the radio data are not shown for clarity. Only data with low index uncertainty, $\Delta \Gamma(3$--$12\, {\rm keV})<0.3$ are plotted.
}
\label{f:X_radio}
\end{figure*}

Figs.\ \ref{f:X_radio}(a--e) illustrate the X-ray spectral variability patterns of Cyg X-1 by showing the relationship between the bolometric flux and the flux in a given band, $F_{\rm X}$. We have fitted these relationships in the hard state (red in Figs.\ \ref{f:lc_bol}--\ref{f:hot_bol}) by a power law, $F_{\rm bol}\propto F_{\rm X}^{p'}$ (see Appendix \ref{fits}). The indices found are given in Table \ref{t:fit}. We see that $p'\simeq 1$ in the 14--150 keV range. This corresponds to a spectrum of constant shape moving up and down in normalization only. On the other hand, the dependencies are substantially weaker than direct proportionality in the 1.5--3 keV and 150--195 keV bands. 

The intermediate state (green in Figs.\ \ref{f:lc_bol}--\ref{f:hot_bol}) shows a different variability pattern. Up to 12 keV, the intermediate state points are quite similar to those in the hard state, though they extend to higher fluxes. However, the dependence changes sign at $E\ga 20$ keV and it becomes increasingly negative (i.e, the local X-ray flux is anti-correlated with $F_{\rm bol}$), as shown in Fig.\ \ref{f:X_radio}(e) for 75--100 keV. 

The soft state (blue in Figs.\ \ref{f:lc_bol}--\ref{f:hot_bol}) shows an approximate dependence of $F_{\rm bol}\simprop F_{\rm X}$ over all the observed energy range. However, the scatter increases with the increasing energy. 

For coronal models (e.g., \citealt{mf02,merloni03}), the flux, $F_{\rm hot}$, due to emission of coronal hot electrons is an important quantity. It is predicted to be constant for coronae above gas-pressure dominated discs, and to decrease as $\dot M^{-1/2}$ for radiation-pressure dominated discs, see a discussion in \citet{mhd03}. It is equal to the bolometric X-ray flux minus the disc blackbody contribution. Ideally, to calculate $F_{\rm hot}$ accurately one should fit each spectrum with a model containing a disc blackbody, and then subtract its total flux from the resulting model bolometric flux. Given our method of determining $F_{\rm bol}$, this is not possible. Instead, we use the flux above an energy approximately dividing the spectral region dominated by the blackbody and that dominated by Comptonization. Specifically, we assume $F_{\rm hot}= F(>\!\!3\,{\rm keV})$ in the soft state, and $=F(>\!\!1.5\,{\rm keV})$ in the intermediate and hard states. These choices are based on the Cyg X-1 spectra shown in \citet{zg04}. Fig.\ \ref{f:hot_bol} shows the resulting estimate of  the coronal emission as a fraction of $F_{\rm bol}$. We see it is approximately constant in the soft state.

\subsection{Radio/X-ray correlation}
\label{corr}

\begin{table*}
\centering
\caption{The values of the fit coefficients (the radio flux normalization, $F_{\rm R,0}$, calculated at $\langle F_{\rm X}\rangle$, and the index, $p$), equation (\ref{fit}), the Spearman's correlation coefficient, $r_{\rm s}$, and its null hypothesis probability, $P_{\rm s}$, for the radio/X-ray correlation. We also give the value of the index, $p'$, for the $F_{\rm bol}$--$F_{\rm X}$ correlation in the hard state, and the $p/p'$ ratio. Only the hard state data have been fitted except for the case of $F({\rm 15\, GHz})$--$F_{\rm hot}$, where the data in all three states have been included. The uncertainties are given at $1\sigma$.
}
\begin{tabular}{cccccccccc}
\hline
Correlation & $F_{\rm R,0}\,$[mJy] &$\langle F_{\rm X}\rangle\,$[keV cm$^{-2}$ s$^{-1}$]  & $p$ & $p'$ & $p/p'$ & $r_{\rm s}$ & $P_{\rm s}$ & $F_{\rm R,min}\,$[mJy] \\
\hline
ASM\\
$F({\rm 2.25\, GHz})$--$F(1.5$--$3\,{\rm keV})$ &  $15.9\pm 0.2$ & 2.09 & $0.75\pm 0.06$ & $0.59\pm 0.01$ & $1.27\pm 0.08$ & 0.26 & $3\times 10^{-13}$ & 10\\
$F({\rm 8.3\, GHz})$--$F(1.5$--$3\,{\rm keV})$ &  $17.0\pm 0.2$ & 2.11 & $0.79\pm 0.06$ &     {\tt "}                          & $1.34\pm 0.07$ & 0.30 & $3\times 10^{-17}$ & 10\\
$F({\rm 15\, GHz})$--$F(1.5$--$3\,{\rm keV})$ &  $12.2\pm 0.1$ & 2.10 & $0.91\pm 0.03$ &  {\tt "}                              & $1.54\pm 0.04$ & 0.49 & $<10^{-20}$ & 5\\
$F({\rm 2.25\, GHz})$--$F(3$--$5\,{\rm keV})$ &  $15.9\pm 0.2$ & 1.52 & $1.07\pm 0.09$ & $0.84\pm 0.01$ & $1.27\pm 0.09$ & 0.25 & $2\times 10^{-12}$ & 10\\
$F({\rm 8.3\, GHz})$--$F(3$--$5\,{\rm keV})$ &  $17.0\pm 0.2$ & 1.53 & $1.13\pm 0.09$ &    {\tt "}                           & $1.35\pm 0.08$ & 0.27 & $1\times 10^{-14}$ & 10\\
$F({\rm 15\, GHz})$--$F(3$--$5\,{\rm keV})$ &  $12.2\pm 0.1$ & 1.54 & $1.34\pm 0.05$ &    {\tt "}                            & $1.59\pm 0.04$ & 0.51 & $<10^{-20}$ & 5\\
$F({\rm 2.25\, GHz})$--$F(5$--$12\,{\rm keV})$ &  $15.9\pm 0.2$ & 3.10 & $1.28\pm 0.12$ & $0.90\pm 0.01$ & $1.42\pm 0.09$ & 0.23 & $6\times 10^{-11}$ & 10\\
$F({\rm 8.3\, GHz})$--$F(5$--$12\,{\rm keV})$ &  $17.0\pm 0.2$ & 3.10 & $1.35\pm 0.12$ &    {\tt "}                           & $1.50\pm 0.09$ & 0.24 & $1\times 10^{-11}$ & 10\\
$F({\rm 15\, GHz})$--$F(5$--$12\,{\rm keV})$ &  $12.2\pm 0.1$ & 3.09 & $1.59\pm 0.06$ &     {\tt "}                          & $1.77\pm 0.04$ & 0.50 & $<10^{-20}$ & 5\\
\hline
BAT\\
$F({\rm 15\, GHz})$--$F(14$--$20\,{\rm keV})$ &  $12.6\pm 0.2$ & 1.75 & $1.68\pm 0.12$ & $1.03\pm 0.01$ & $1.63\pm 0.07$ & 0.56 & $<10^{-20}$ & 5\\
$F({\rm 15\, GHz})$--$F(20$--$24\,{\rm keV})$ &  $12.6\pm 0.2$ & 1.00 & $1.75\pm 0.13$ & $1.04\pm 0.01$ & $1.68\pm 0.07$ & 0.56 & $<10^{-20}$ & 5\\
$F({\rm 15\, GHz})$--$F(24$--$35\,{\rm keV})$ &  $12.6\pm 0.2$ & 2.30 & $1.78\pm 0.13$ & $1.05\pm 0.01$ & $1.69\pm 0.07$ & 0.53 & $<10^{-20}$ & 5\\
$F({\rm 15\, GHz})$--$F(35$--$50\,{\rm keV})$ &  $12.6\pm 0.2$ & 2.47 & $1.78\pm 0.15$ & $1.05\pm 0.01$ & $1.69\pm 0.08$ & 0.47 & $2\times 10^{-20}$ & 5\\
$F({\rm 15\, GHz})$--$F(50$--$75\,{\rm keV})$ &  $12.6\pm 0.2$ & 3.04 & $1.77\pm 0.16$ & $1.05\pm 0.01$ & $1.69\pm 0.09$ & 0.42 & $4\times 10^{-16}$ & 5\\
$F({\rm 15\, GHz})$--$F(75$--$100\,{\rm keV})$ &  $12.6\pm 0.3$ & 2.22 &$1.74\pm 0.19$ & $1.04\pm 0.02$ & $1.67\pm 0.11$ & 0.33  & $2\times 10^{-10}$ & 5\\
$F({\rm 15\, GHz})$--$F(100$--$150\,{\rm keV})$ & $12.6\pm 0.3$ & 2.92 &$1.64\pm 0.23$ & $0.99\pm 0.02$ & $1.66\pm 0.14$ & 0.24  & $4\times 10^{-6}$ & 5\\
$F({\rm 15\, GHz})$--$F(150$--$195\,{\rm keV})$ &  $12.6\pm 0.3$ & 1.63 &$0.87\pm 0.18$ & $0.48\pm 0.02$ & $1.81\pm 0.21$ & 0.12 & $0.03$ & 5\\
\hline
ASM+BAT\\
$F({\rm 15\, GHz})$--$F_{\rm bol}$ &  $12.6\pm 0.2$ & 32.6 & $1.68\pm 0.11$ & 1 & -- & 0.58 & $<10^{-20}$ & 5\\
$F({\rm 15\, GHz})$--$F_{\rm hot}$ &  $11.4\pm 0.2$ & 26.1 & $1.73\pm 0.07$ & -- & -- & 0.55 & $<10^{-20}$ & 1\\
\hline
ASM+BATSE\\
$F({\rm 2.25\, GHz})$--$F_{\rm bol}$ &  $15.8\pm 0.2$ & 30.7 & $1.49\pm 0.15$ & 1 & -- & 0.24 & $2\times 10^{-9}$& 10\\
$F({\rm 8.3\, GHz})$--$F_{\rm bol}$ &  $17.0\pm 0.2$ & 30.6 & $1.58\pm 0.18$ & 1 & -- & 0.19 & $9\times 10^{-7}$ & 10 \\
$F({\rm 15\, GHz})$--$F_{\rm bol}$ &  $12.2\pm 0.2$ & 30.2 & $1.81\pm 0.11$ & 1 & -- & 0.46 & $<10^{-20}$ & 5\\
\hline
\end{tabular}
\label{t:fit}
\end{table*}

We now consider correlations between the X-ray and radio fluxes. Figs.\ \ref{f:X_radio}(f--j) shows the relationships between the X-ray flux in selected ASM and BAT bands and the 15 GHz radio flux. The colours identify the spectral state. The reason that there are fewer  points in the BAT range (i--j) is because each requires simultaneous BAT, 15 GHz radio and ASM coverage (the last to determine the spectral state). 

It can be seen that the hard-state fluxes (shown in red) approximately follow a power-law relationship,  $F({\rm 15\,GHz})\propto F_{\rm X}^{p}$, but the power-law index is a strong function of the X-ray energy band. The fitted  parameters  (see Appendix \ref{fits} for the fitting method) are given in Table \ref{t:fit}. The high significance of the correlations is confirmed by the values of the Spearman's correlation coefficient, $r_{\rm s}$, and the null hypothesis probability, $P_{\rm s}$, also given in Table \ref{t:fit}. The index, $p$, changes from $\simeq 0.9$ at 1.5--3 keV to $\simeq 1.7$--1.8 in the 14--150 keV range.

\begin{figure}
\centerline{\includegraphics[width=7cm]{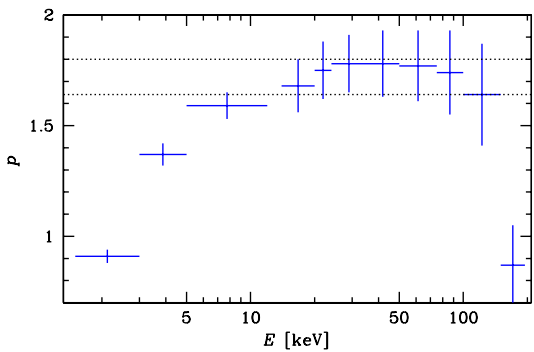}} 
\caption{The power-law index of the hard-state correlation between $F(15\,{\rm GHz})$ and $F_{\rm X}$ as a function of the photon energy, see Table \ref{t:fit}. The dotted horizontal lines show the uncertainty range of $p$ for the Comptonization flux, $F_{\rm hot}$. 
}
\label{f:index}
\end{figure}

The variation of  $p$ with energy is summarised in Fig.\ \ref{f:index}. The low value of the index in the 150--195 keV band  index is apparently due the fact that this energy range corresponds to where the high-energy cut-off of the spectra starts to have a major effect. Thus much of the variability is associated with a different physical process -- changes in the cut-off energy, apparently due to a variable electron temperature.

\begin{figure}
\centerline{\includegraphics[height=6.3cm]{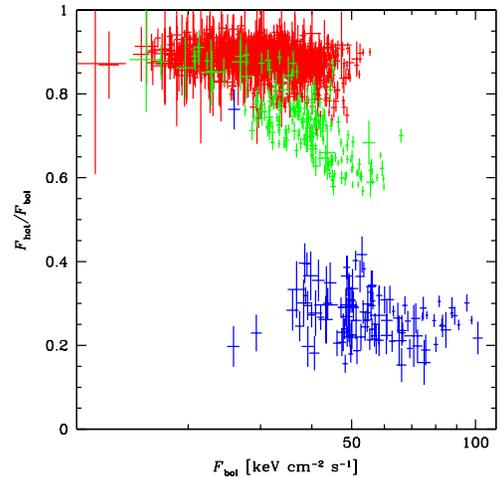}} 
\caption{The flux in the Comptonization part of the spectrum (approximated as described in Section \ref{xray}) divided by the bolometric flux, based on the ASM+BAT data. The red, green and blue symbols correspond to $\Gamma(3$--$12\, {\rm keV}) <1.9$ (hard state), $1.9<\Gamma<2.3$ (intermediate state) and $\Gamma>2.3$ (soft state), respectively.
}
\label{f:hot_bol}
\end{figure}

The relationship between the radio flux, $F(\rm 15\, GHz)$, and the X-ray hardness is illustrated in Fig.\ \ref{f:gamma}(b).  There is a strong correlation in the hard state, $r_{\rm s} = 0.22$, $P_{\rm s} = 2\times 10^{-19}$, which continues up to $\Gamma\sim 2.1$. When the X-ray spectra are soft, there is a strong decrease in the radio emission; for $\Gamma>2.3$ the  (anti-)correlation has $r_{\rm s} = -0.57$, $P_{\rm s} = 2\times 10^{-37}$.

\begin{figure*}
\centerline{\includegraphics[height=8cm]{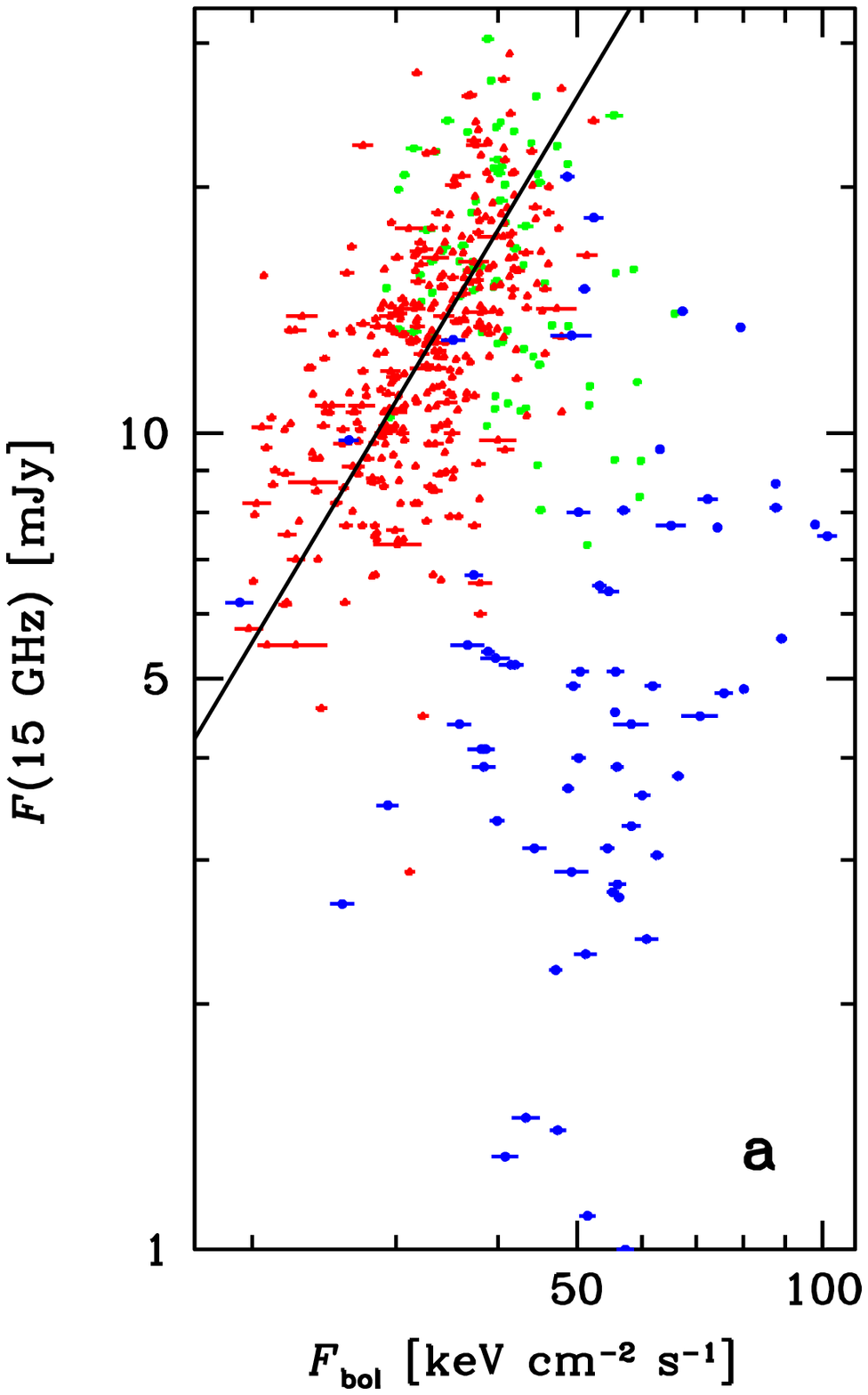}
\includegraphics[height=8cm]{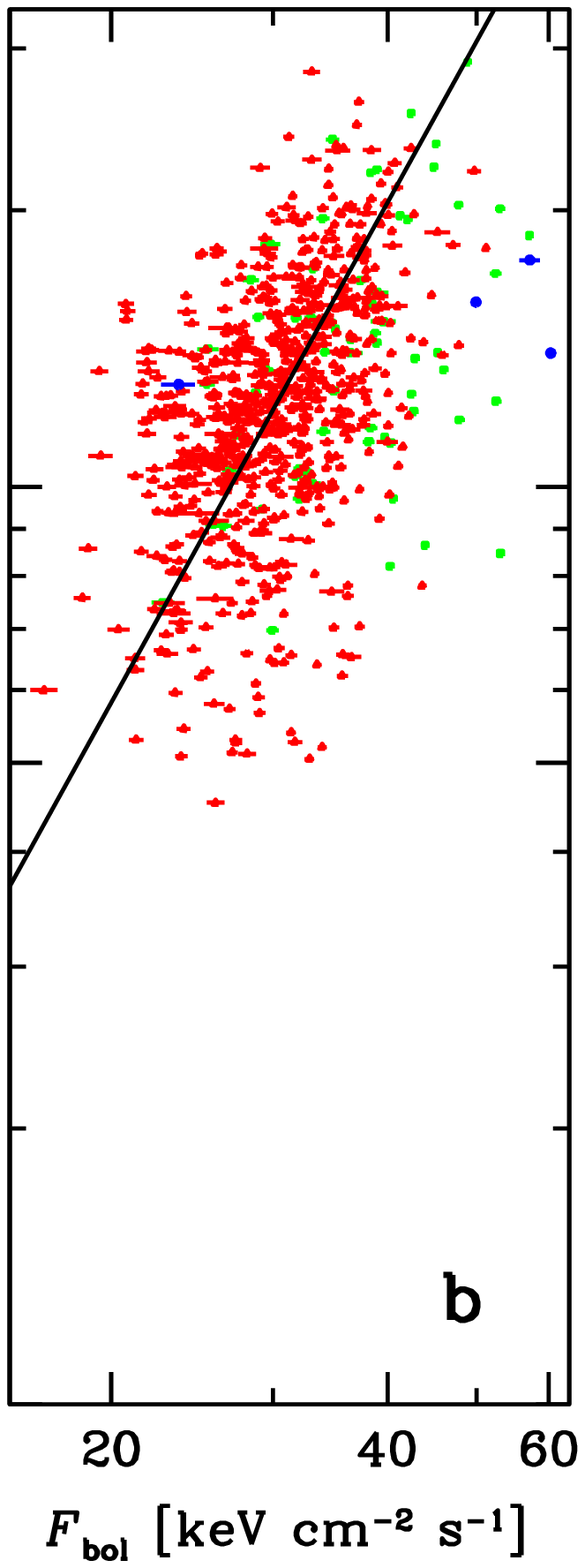}
\includegraphics[height=8cm]{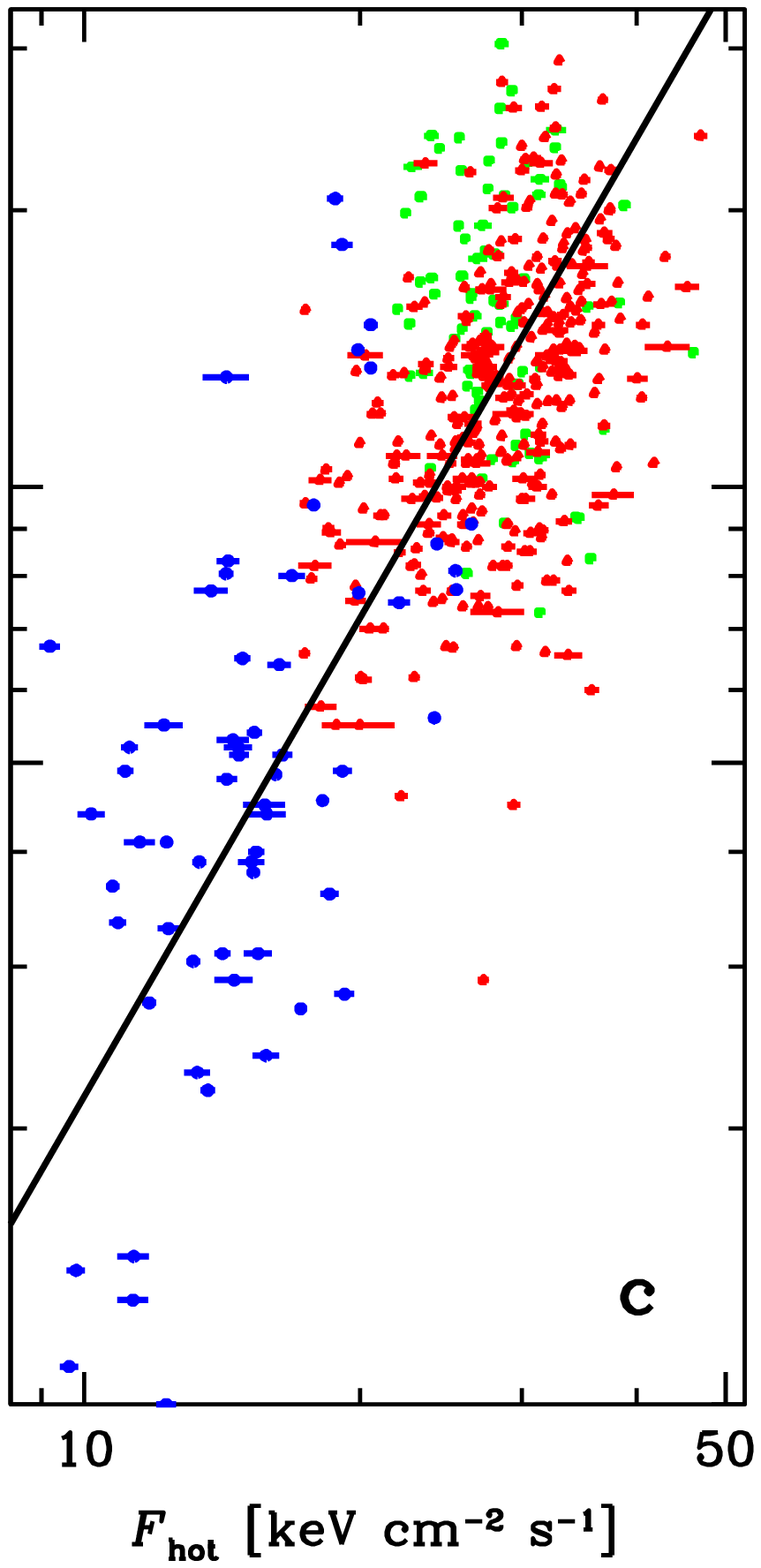}}
\caption{Relationship between the daily-averaged radio flux at 15 GHz and the corresponding (a) $F_{\rm bol}$ from the ASM+BAT data, (b) $F_{\rm bol}$ from the ASM+BATSE data, and (c) the Comptonization flux, $F_{\rm hot}$ from the ASM+BAT data. $F_{\rm hot}$ is approximated as $F(>3\,{\rm keV})$ in the soft state, and $F(>1.5\,{\rm keV})$ in the intermediate and hard states. The red triangles, green squares and blue circles correspond to $\Gamma(3$--12 keV) of $\Gamma<1.9$, $1.9<\Gamma<2.3$ and $\Gamma>2.3$, respectively. The error bars on the radio measurements are not shown for clarity. The solid lines show the fits to the hard-state data for (a--b) and to the data in all states for (c) using equation (\ref{fit}), see Table \ref{t:fit}.
}
\label{f:bol_radio15}
\end{figure*}

Fig.\ \ref{f:bol_radio15}(a--b) shows the relationship between the 15-GHz flux and the bolometric flux. The approximate power-law relationship in the hard state now extends to most of the intermediate state, though it breaks down in the soft state. The fit results for the hard state are given in the third part of Table \ref{t:fit}. Although the low energy data are included in $F_{\rm bol}$, the correlation is much steeper ($p_{\rm bol}=1.68\pm 0.11$) than at $E\leq 5$ keV, where the power law index was $p\simeq 0.9$--1.3. 

As $p_{\rm bol} = {\rm d} \ln F({\rm 15\,GHz}) / {\rm d} \ln F_{\rm bol}$, while  $p = {\rm d} \ln F({\rm 15\, GHz}) / {\rm d} \ln F_{\rm X}$ and $p' = {\rm d} \ln F_{\rm bol} / {\rm d} \ln F_{\rm X}$ (where the differential coefficients are those of the fitted functional dependencies), in the absence of  hidden systematic variability patterns one might expect $p_{\rm bol}\simeq p/p'$, and indeed it can be seen from Table \ref{t:fit} that this is very close to being the case, especially at $E\ga 5$ keV.

We note that Figs.\ \ref{f:X_radio}(f--j) show that the radio flux in the states with a strong X-ray blackbody contributions, i.e., intermediate and soft, depends as a power law on the emission at high energies, $\ga 5$ keV. This indicates that the level of the radio emission is associated with the high-energy electrons in the X-ray emitting region, presumably a corona in the intermediate and soft states (e.g., \citealt{gierlinski99}). To test this hypothesis, we have considered the dependence of the 15 GHz flux on the integrated X-ray flux integrated excluding the disc blackbody contribution, i.e., $F_{\rm hot}$ calculated in Section \ref{xray}. Fig.\ \ref{f:bol_radio15}(c) shows the resulting relationship. In contrast to Fig.\ \ref{f:bol_radio15}(a--b), we see that now there is an approximate single power-law dependence encompassing {\it all\/} the states of Cyg X-1, with $p=1.73\pm 0.07$. 

\begin{figure}
\centerline{\includegraphics[height=7cm]{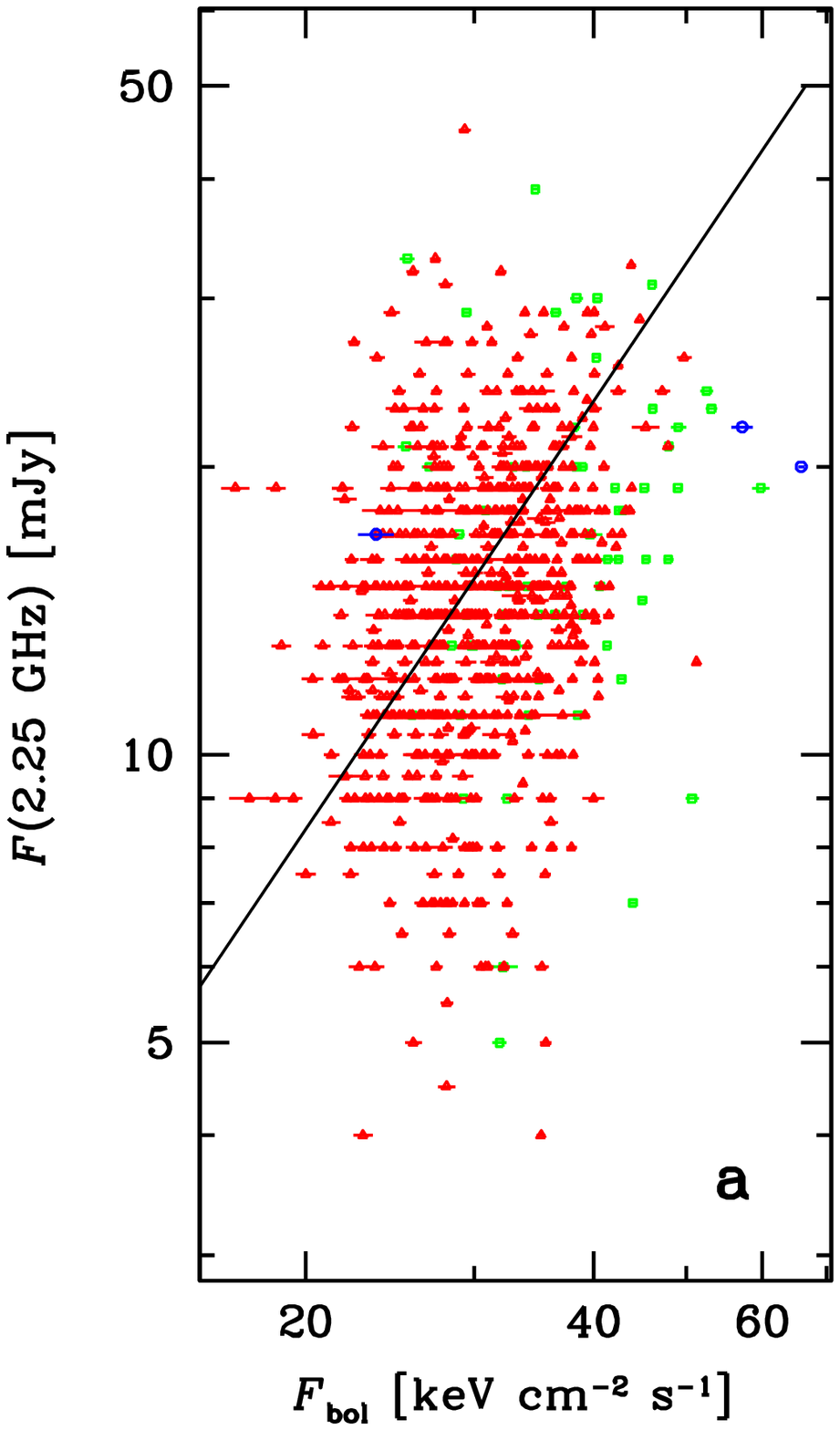}
\includegraphics[height=7cm]{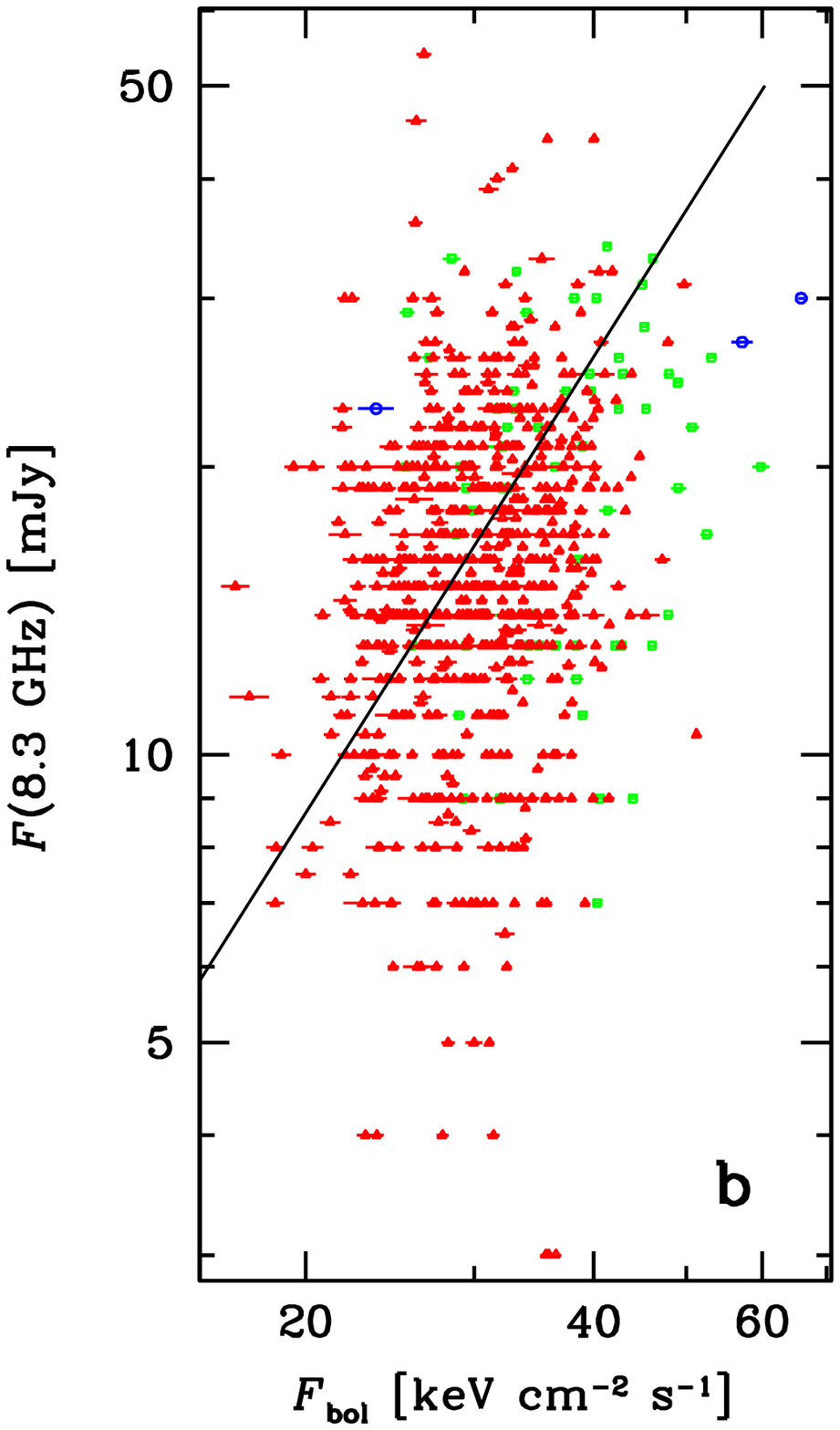}}
\caption{Correlations of the daily-averaged radio flux at (a) 2.25 GHz and (b) 8.3 GHz with the corresponding bolometric flux estimated based on the ASM and BATSE data. The red triangles, green squares and blue circles correspond $\Gamma(3$--12 keV) of $<1.9$, $1.9<\Gamma<2.3$ and $>2.3$, respectively. The error bars on the radio measurements are not shown for clarity. The solid lines show the best fits of the power-law correlation to the hard-state data, see Table \ref{t:fit}. Only the data above 10 mJy have been fitted.}
\label{f:radio28}
\end{figure}

The GBI data provide information at frequencies below 15 GHz and were simultaneous with part of the ASM and BATSE data. We find correlations at 2.25 and 8.3 GHz in the hard state similar to that found at 15 GHz -- see Fig.\ \ref{f:radio28} and the first and last sections of Table \ref{t:fit}. Despite the relatively low quality of those data, the correlations are highly significant, as seen from  the low values of $P_{\rm s}$. The fitted indices are, however,  subject to some systematic errors associated with the choice of the minimum radio flux, $F_{\rm R,min}$, considered. Using $F_{\rm R,min}= 15$ mJy at 2.25 GHz and 8.3 GHz would lead to somewhat lower values of $p$ than with the limit of 10 mJy adopted here. In any case, $p$ decreases with the decreasing frequency.

Fig.\ \ref{f:sprx} shows the broad-band spectra of Cyg X-1 from radio to X-rays. The black circles and vertical error bars show the average hard-state radio fluxes from \citet{fender00}. The dotted line shows an $\alpha=0$ ($F_E\propto E^{-\alpha}$, $\alpha=\Gamma-1$) dependence, which approximately fits the hard-state radio data. We show the line extending to 0.1 eV, which approximately correspond to the turnover frequency (at which the jet becomes optically thin to synchrotron self-absorption) of $2.9\times 10^{13}$ Hz found by \citet{rahoui11}. The red, green and blue crosses show the (geometric) average fluxes in the hard, intermediate and soft state, respectively, in the radio and X-ray ranges. Only the ASM and BAT data simultaneous with the 15 GHz monitoring were included. The 2.25 GHz and 8.3 GHz data are the averages over all the data for which there were simultaneous ASM data for determining the spectral state. There are no blue crosses at those frequencies because there was virtually no soft state during the GBI monitoring. The cyan crosses and error bars show the measurements of \citet{persi80} and \citet{mirabel96} of infrared emission, which is from the companion and its wind. The dashed lines are extrapolations of the 20--35 keV average spectra (which emission contains virtually no contribution from the disc blackbody) down to 0.1 eV.

\section{Discussion}
\label{discussion}

\subsection{The nature of the X-ray variability}
\label{xvar}

We find that the hard state, with $\Gamma(3$--$12\,{\rm keV})<1.9$, has the dominant variability pattern of the $\sim$14--150 keV spectrum with a constant shape moving up and down, see Figs.\ \ref{f:X_radio}(c--e), \ref{f:gamma}(a). $F_{\rm bol}$ changes by a factor of $\simeq$5 within the hard state. This pattern can be caused by a changing accretion rate, $\dot M$, in a stable geometry (Z02). 

The superorbital variability has been found by ZPS11 to show the same variability pattern. However, its amplitude is only by about $\pm 0.2$ above 15 keV (ZPS11), and thus it can explain only a small part of the observed factor of $\sim$5 variability.

On the other hand, the softest ($<$3 keV) and hardest ($>$150 keV) X-rays have faster, roughly quadratic,  relationships with $F_{\rm bol}$. Below 3 keV, absorption is strong and there is a soft-excess component (e.g., \citealt{gierlinski97}). The observed relationship is likely to be connected to the absorbing column increasing at decreasing fluxes (whereas $F_{\rm X}$ have been calculated under the assumption of a constant absorbing column, see Appendix \ref{bol}). Also, it is possible that the soft excess is weaker at lower fluxes. We cannot evaluate the relative contributions of these effects based on our data alone. 

We note here that ZPS11 argued for precession of the accretion disc as the preferred model for the superorbital variability based on the constancy of the photon index above 20 keV during the superorbital cycle. In the light of the finding of this work, precession is still possible but not proven. Our results in Figs.\ \ref{f:X_radio}(a--e) show $F_{\rm bol}\propto F_{\rm X}$ in the 14--150 keV range, which is compatible with the spectrum with the same photon index moving up and down. In the present case, the amplitude of this variability is by a factor of $\simeq$5, and it has to be driven by changing $\dot M$. Also, Figs.\ \ref{f:X_radio}(a--b) are compatible with increasing absorption at low fluxes, similar to the behaviour found during the superorbital cycle (ZPS11). Thus, it is possible that the superorbital variability of Cyg X-1 is driven by a modulation of $\dot M$ on the time scales of about 150 or 300 d, which cause, however, would be unknown. 

Above 150 keV, the hard-state spectra show a cut-off, which in the framework of the thermal Comptonization model is expected at $E\ga (2$--$3) kT_{\rm e}$, where $T_{\rm e}$ is the temperature of the Comptonizing electrons.  The low value of the correlation index, $p'$, at these energies indicates that $kT_{\rm e}$ increases with increasing $F_{\rm bol}$. This trend is opposite to that modeled by Z02, where the increased plasma cooling at a higher luminosity leads to a small decrease of $kT_{\rm e}$ (see fig.\ 14c of Z02). We note, however, that given the strong noise in the 150--195 keV data, the exact variability pattern in this band is relatively uncertain. Also, the trend of fig.\ 14(c) Z02 may be overcome by the increase of $kT_{\rm e}$ for harder spectra, see figs.\ 14(a--b) in Z02, and see fig.\ 13 in Z02 for example spectra from pointed observations showing this trend.  

For the intermediate state, up to $\Gamma\simeq 2.3$--2.5, the bolometric flux increases with the increasing $\Gamma$, see Fig.\ \ref{f:gamma}(a). The broad-band variability shows also a pivot at $\sim$20 keV, above which the intermediate-state X-ray fluxes in Figs.\ \ref{f:X_radio}(a--e) change from being correlated with $F_{\rm bol}$ to an anti-correlation. This has been modelled by Z02 and \citet*{gzd11} as due to a changing ratio of the flux irradiating a Comptonizing plasma to that dissipated in that plasma (see figs.\ 14a--b in Z02, 10c in \citealt{gzd11}). This may be caused by a varying the inner radius of an accretion disc surrounding the hot plasma \citep*{dgk07}.

The soft-state X-ray bolometric flux is dominated by photons below $E\sim$3 keV. Below this energy the soft-state spectra are dominated by the disc component and above it, by a high-energy tail. The soft state variability has been modelled by \citet*{cgr01}, and later by Z02 and \citet{gzd11} as a variable non-thermal corona above a stable disc, see, e.g., fig.\ 16 in Z02 and fig.\ 10c in \citet{gzd11}. However, the disc blackbody emission is stable only on short time scales, whereas it does change significantly on long time scales, see, e.g., Fig.\ \ref{f:X_radio}(a). The resulting variability pattern is of the overall spectrum changing its normalization but with much more scatter if $F_{\rm X}$ vs.\ $F_{\rm bol}$ at high energies, dominated by the emission of the fast-varying corona, than in the blackbody component.

\begin{figure*}
\centerline{\includegraphics[height=9cm]{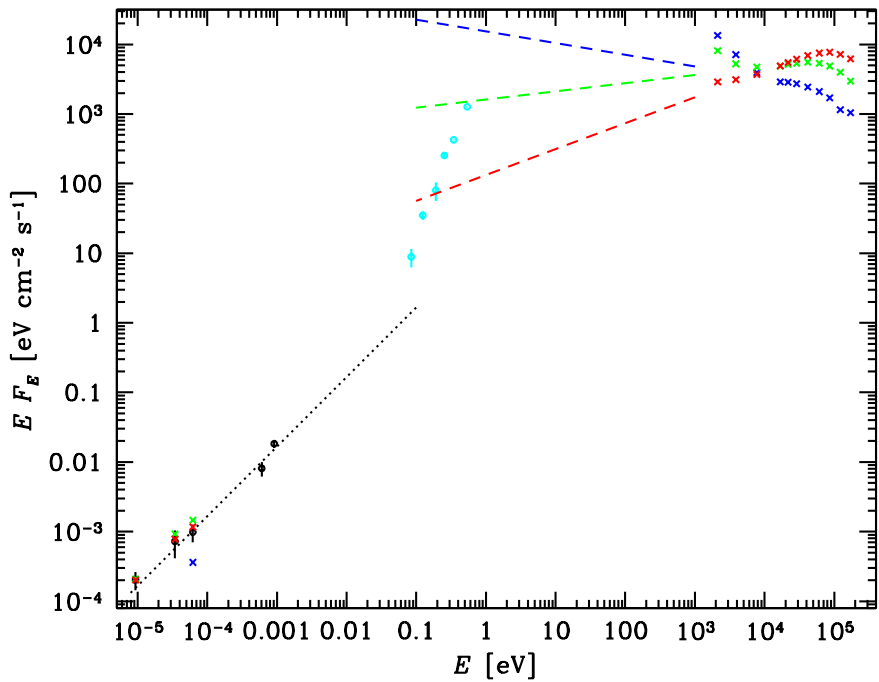}}
\caption{The broad-band spectra of Cyg X-1 from radio to X-rays. The black circles and vertical error bars show the average hard-state radio fluxes from \citet{fender00}, and the dotted line shows the $F_\nu\propto \nu^0$ dependence fitting those points extended up to the turnover energy of 0.1 eV. The red, green and blue crosses show our results in the hard, intermediate and soft state, respectively. The X-ray data are absorption-corrected. The cyan circles and error bars show the infrared fluxes, which are mostly from emission of the companion and its wind. The dashed lines are extrapolations of the 20--35 keV average spectra down to 0.1 eV.
}
\label{f:sprx}
\end{figure*}

Figs.\ \ref{f:gamma}(a), \ref{f:lc_bol} show that the daily-averaged $F_{\rm bol}$ changes, over the three states, by a factor of $\simeq$10. As expected, the highest and lowest fluxes correspond to the soft and hard state, respectively, consistent with the view that state transitions in Cyg X-1 are driven by changes of the accretion rate. The soft-state $F_{\rm bol}$ is generally about a factor of $\simeq$3 higher that in the hard state.  We note, however, that $F_{\rm bol}$ in the intermediate and soft states can become less than that in the hard state, see Figs.\ \ref{f:gamma}(a), \ref{f:lc_bol}, \ref{f:X_radio}(a--e). This appears to be the first such finding in Cyg X-1. It may be indicative of hysteresis, usually present in transient sources (e.g., \citealt{dunn08}). Its presence there is seen as a characteristic q track in the flux/hardness diagram, see, e.g., fig.\ 1 in \citet{russell10}. An analogue to it is visible in Fig.\ \ref{f:gamma}(a), where some points belonging to the soft and intermediate  states have $F_{\rm bol}$ below that of the hard state. However, the tracks covered are much more chaotic than those in transients, where the hard-to-soft transitions occur at clearly higher flux than the reverse ones. Although we consider this result for Cyg X-1 secure, it relies on the absorption-corrected ASM observations, and should be tested by pointed X-ray observations. 

\subsection{The nature of the radio/X-ray correlation}
\label{rx}

Our findings concerning radio/X-ray correlations fall into two categories. Those regarding the dependence of the correlation on the X-ray energy range and on the spectral state are basically independent of the uncertain strength of free-free absorption in the stellar wind from the donor of Cyg X-1. On the other hand, the values of the radio/X-ray correlation indices are, most likely, affected by the free-free absorption. 

\subsubsection{The correlation dependencies on the X-ray energy range and on the spectral state}
\label{range}

One of our findings is the strong dependence of the radio/X-ray correlation index on the energy band, see Table \ref{t:fit} and Fig.\ \ \ref{f:index}. The energy-dependence of the correlation can be understood as an intrinsic dependence of the radio flux on the bolometric flux combined with flux-dependent changes in the spectral shape (seen in our data at $E<12$ keV and $E>150$ keV). The latter leads to the bolometric flux and that in a narrow X-ray band being non-proportional in a way that depends on the band chosen.

This dependence has mostly been ignored in previous work on this subject except, e.g., \citet{mhd03}, where its effect has been accounted for by a theoretical model, \citet{z04}, who fitted the radio correlation with $F_{\rm bol}$ in GX 339--4, obtaining $p_{\rm bol}=0.79\pm 0.07$, and \citet{corbel03}, who calculated the correlation in GX 339--4 in four separate bands between 3 keV and 200 keV, but have not calculated the correlation index for $F_{\rm bol}$. 

In models in which the accretion rate, $\dot M$, governs both the accretion flow and the jet (see Section \ref{accretion} below), it is crucial to determine its value. It is related to the bolometric flux by,
\begin{equation}
\dot M= {4\upi d^2 F_{\rm bol}\over \epsilon c^2},
\label{mdot}
\end{equation}
where $d$ is the source distance, and $\epsilon$ is the accretion radiative efficiency. Our results indicate that the jet power is a fraction not of the total accretion power but only of that responsible for emission associated with hot electrons in the accretion flow, $F_{\rm hot}$. In this case we should write $\dot M \propto F_{\rm hot}/\epsilon_{\rm hot}$, where $\epsilon_{\rm hot}$ is the efficiency of producing the hot flow. However, the difference between the two is negligible in the hard state, where the hot flow is energetically dominant. 

The common procedure has been to use a narrow band, such as 3--9 keV (e.g., \citealt{coriat11}), as the proxy for $F_{\rm bol}$, i.e. implicitly assuming $F_{\rm X}\propto F_{\rm bol}$, and then proceed to theoretical interpretation. However, we have shown that this can lead to significant errors. For our data, we can estimate the 3--9 keV flux as the sum of the 3--5 keV flux and a fraction of the 5--12 keV flux, the fraction being determined assuming a power law with the 3--12 keV photon index. Using the flux in this band, we find $p=1.17\pm 0.10$, $1.23\pm 0.10$, and $1.48\pm 0.05$, for 2.25 GHz, 8.3 GHz, and 15 GHz, respectively. Comparing with Table \ref{t:fit}, we see that the 3--9 keV values of $p$ are smaller than those for $p_{\rm bol}$ by $\simeq 0.2$--0.3. This problem occurs for any system with spectral variability, which is generally the case in either black-hole or neutron-star X-ray binaries. 

Our next main finding is that we are able to extend the correlation to the intermediate and soft states. The radio emission has been known to correlate in the soft states of some objects with the level of the high-energy tail, e.g., in Cyg X-3 \citep{szm08}; however, that dependence was different from that in the hard state. Here, we find that if we consider, instead of the bolometric flux, that in the emission in the hot plasma only (i.e., $F_{\rm bol}$ minus the disc blackbody component), we obtain a single radio/X-ray power law correlation going through all the states. Its index, $1.73\pm 0.07$, is consistent within uncertainties with the correlation index for $F_{\rm bol}$ (for the ASM+BAT data) in the hard state only, $1.68\pm 0.11$. This appears to indicate that the jet is launched from the hot electrons responsible for the observed X-ray emission rather than by the optically-thick disc, and that the launching mechanism may be similar in the hard spectral state and in the soft one. We stress that Cyg X-1 appears to always have some level of the high energy tail in the soft state, and consequently has often relatively strong radio emission in that state, as also found by \citet{rushton11}. This is unlike low-mass black-hole systems, which often go to an extended soft state in which the tail virtually disappears (e.g., \citealt{gd04}), and where consequently there is virtually no radio emission. Also,  Cyg X-1 does not show the very high state, where ejection of radio-emitting blobs occurs instead of a jet, and where the launching mechanism is likely to be different. No such blob ejections have indeed been found during the 2010 state transition by \citet{rushton11}.

We have also found that the maximum of the radio emission occurs for relatively soft X-ray spectra, with $\Gamma(3$--$12\,{\rm keV})\simeq 2$--2.1. Thus, the transition to the soft state is first associated with an increase of the radio emission and only then with a decrease.  

\subsubsection{Free-free absorption and the intrinsic correlation index}
\label{free-free}

\begin{figure*}
\centerline{\includegraphics[width=11cm]{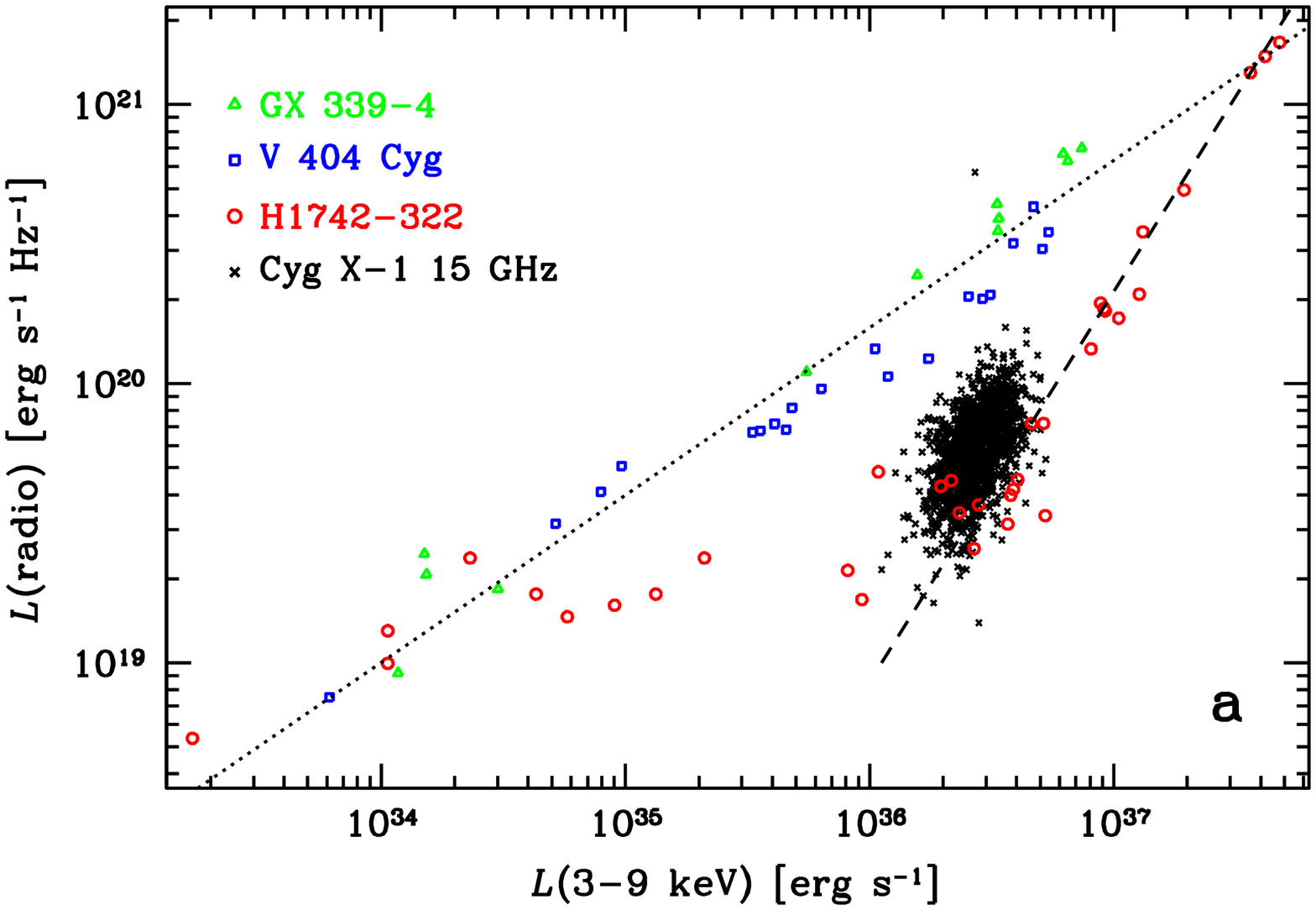}}
\centerline{\hbox to 0.5cm{\hfill} \includegraphics[width=5.5cm]{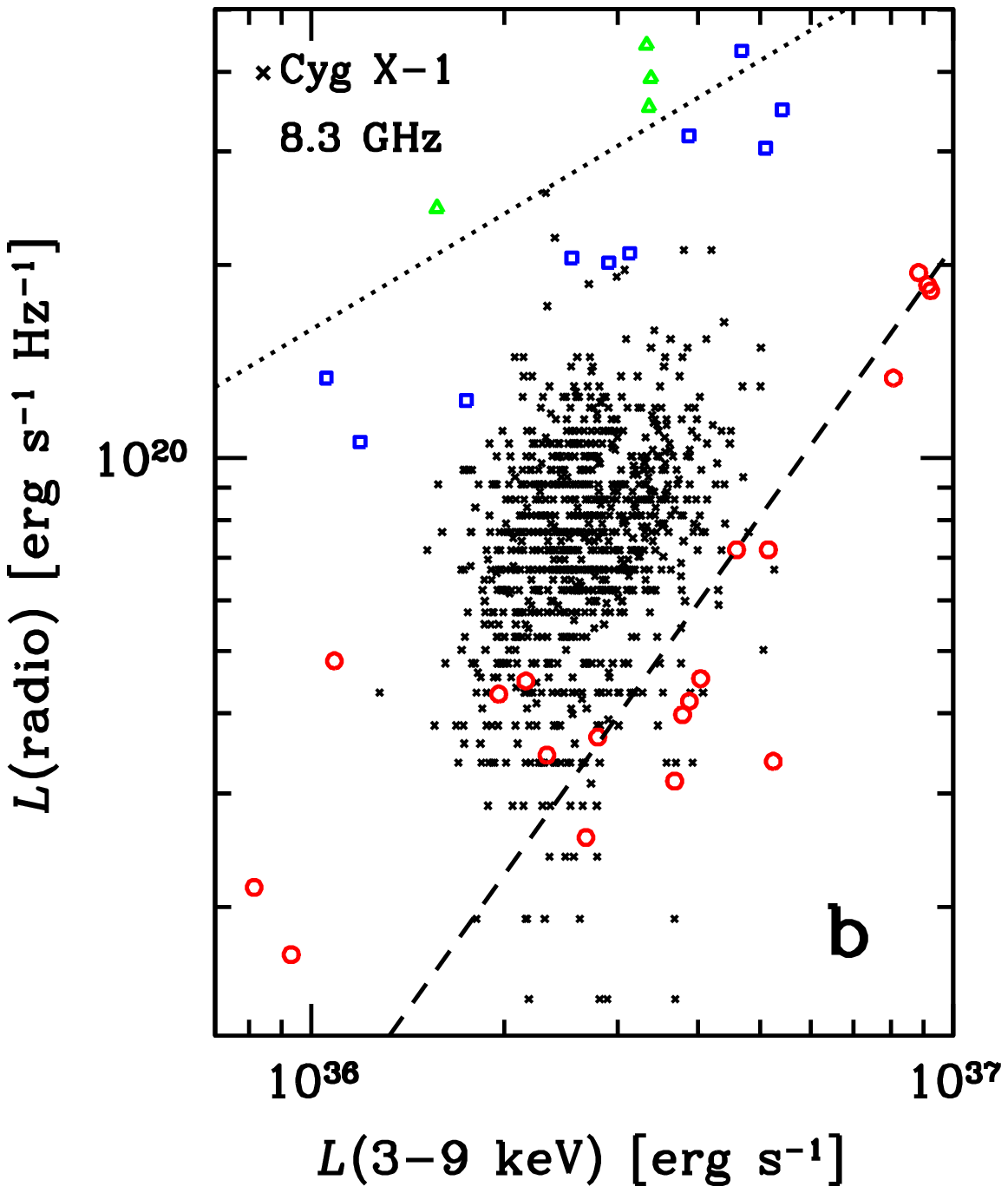} \includegraphics[width=5.5cm]{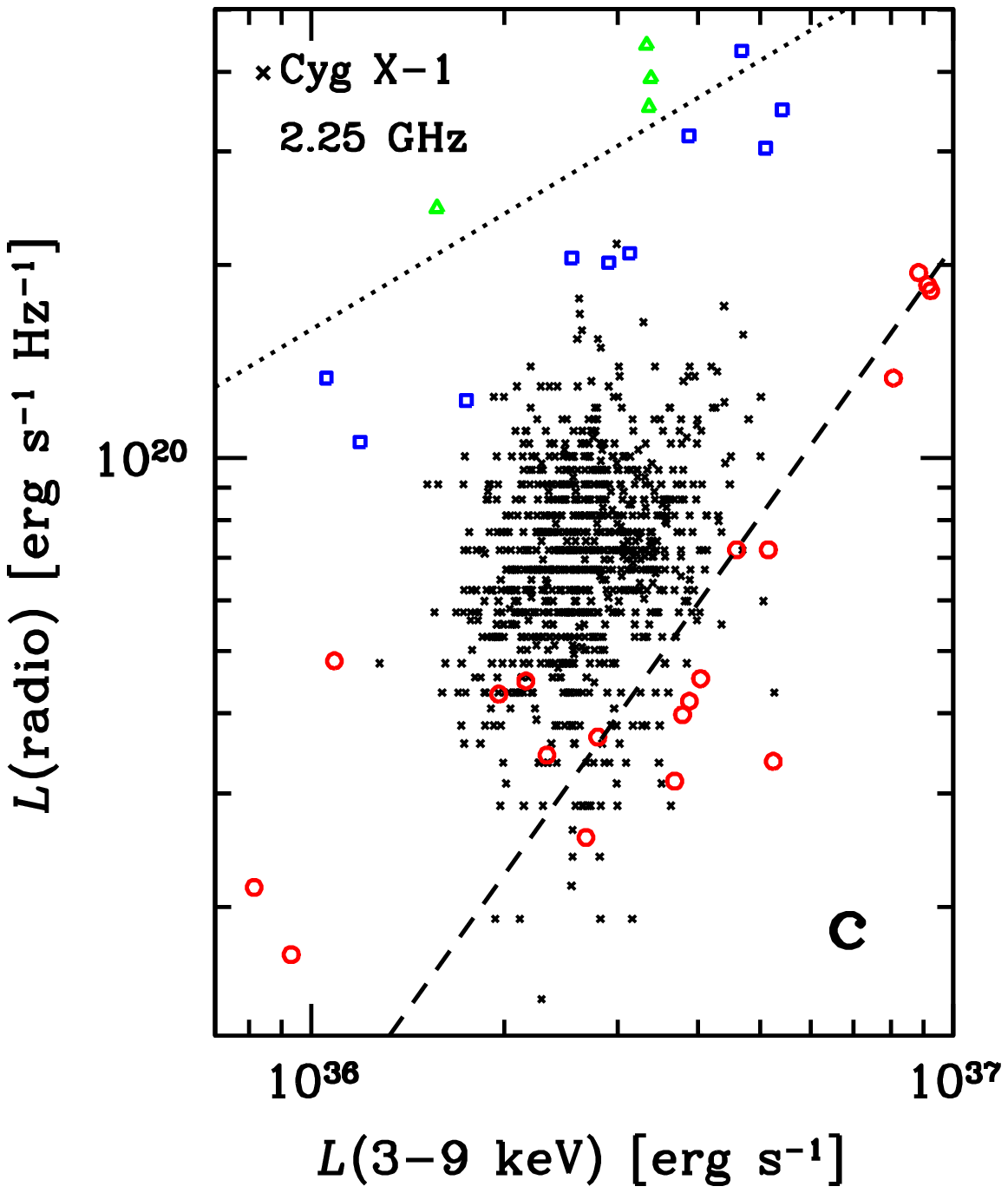}}
\caption{Comparison of the radio/X-ray correlation for Cyg X-1 in the hard state (assuming $d=2$ kpc; black crosses) with those for GX 339--4 (\citealt{corbel03}; green crosses), V404 Cyg (\citealt{corbel08}; blue squares) and H1743--322 (\citet{coriat11}; red circles). The dotted and dashed lines have $p=0.6$ and $p=1.4$, respectively, approximately fitting the two apparent branches of the correlation for the three sources. The Cyg X-1 radio data are for (a) 15 GHz, (b) 8.3 GHz, and (c) 2.25 GHz. 
}
\label{f:correlation}
\end{figure*}

Radio emission of Cyg X-1 from 2.25 GHz to 15 GHz is known to exhibit orbital modulation due to free-free absorption in the stellar wind of the companion \citep{pfb99}. The full modulation depth changes from $\sim$4 per cent at 2.25 GHz to 30 per cent at 15 GHz \citep{l06}. Thus, free-free absorption certainly affects the observed radio fluxes. However, it is not trivial to determine the degree of the attenuation, $\exp(-\bar{\tau}_{\rm ff})$, where $\bar{\tau}_{\rm ff}$ is the orbit-averaged optical depth. Using an estimate of the mass loss rate from the companion towards the polar regions (where the jet radio emission is absorbed) and of the inclination, and calculating the temperature of the wind irradiated both by the star and the X-ray source, we can use the strength of the observed orbital modulation to calculate the height along the jet, $z$, from which the bulk  of the radio emission at a given frequency is received \citep{sz07}. This height corresponds to the place above which the jet becomes optically thin to synchrotron self-absorption \citep{bk79}. Based on the value of $z$, we can calculate the optical depth, $\bar{\tau}_{\rm ff}$, through the wind, provided its density and temperature profile are known. 

Given that a radio flux, $F_{\rm R}=F_{\rm R,intr} \exp(-\bar{\tau}_{\rm ff})$ and $p={\rm d}\ln F_{\rm R}/{\rm d}\ln F_{\rm X}$, where $F_{\rm X}$ is now either the flux in an X-ray band, the bolometric flux or the flux emitted by hot electrons, an index, $p$, derived from the observations is related to the corresponding intrinsic index, $p_{\rm intr}={\rm d}\ln F_{\rm R,intr}/{\rm d}\ln F_{\rm X}$, by
\begin{equation}
p_{\rm intr}= r p,\quad r= {{\rm d} \ln F_{\rm R,intr}\over {\rm d}\ln F_{\rm R}}
=1+{{\rm d}\bar{\tau}_{\rm ff} \over {\rm d}\ln F_{\rm R}}.
\label{index}
\end{equation}
Generally, we expect the jet power to positively correlate with the height along the jet where the emission becomes optically thin to synchrotron self-absorption. Thus $F_{\rm R}$ should be anti-correlated with the depth within the stellar wind, given by $\bar\tau_{\rm ff}$, which then implies $r<1$. Consequently, the effect of free-free absorption of the radio fluxes is to increase their observed variability range with respect to the intrinsic range, and to make the radio/X-ray correlation steeper. This effect has been suggested to qualitatively explain the relative steepness of the correlation in Cyg X-1 by \citet{gfp03}.

A quantitative investigation of this effect has been made by \citet{zdz11}, where it was found that the depth of the orbital radio modulation at 15 GHz (related to $\bar\tau_{\rm ff}$) indeed significantly anti-correlates with $F_{\rm R}$. This can yield the factor $r$ of equation (\ref{index}), provided the wind parameters are specified. Here, an important factor is the weakness of the polar component of the wind of Cyg X-1 recently found by \citet{gies08}. Consequently, \citet{zdz11} found values of $\bar{\tau}_{\rm ff}$ significantly lower than those obtained by \citet{sz07}, who assumed the wind is isotropic. The best resulting estimate of $r$ is $\sim$0.8 at 15 GHz, which yields $p_{\rm bol,intr}\simeq 1.3\pm 0.1$ for the correlation with the bolometric flux estimated based on the ASM+BAT data. That estimate also implies that the wind attenuation of the 15 GHz flux is rather small, which is consistent with the observed 2--220 GHz spectrum (shown also in Fig.\ \ref{f:sprx}) being relatively straight with $\alpha\simeq 0$, and not showing deviations from that power law by more a factor of $\sim$2. Also, the relative mildness of the wind attenuation is consistent with the orbital modulation at 89 GHz, 146 GHz and 221 GHz not being reported by \citet{fender00}. We caution, however, that the theoretical model of \citet{zdz11} is a subject to uncertainties due to the uncertain wind density in polar regions. 

In order to show the Cyg X-1 hard-state correlation in the context of other sources, we have compared it to those of the black-hole binaries H1743--322, GX 339--4 and V404 Cyg, see fig.\ 5 in \citet{coriat11}. We have calculated the 3--9 keV fluxes as described in Section \ref{range} above. To obtain the luminosities, we have assumed isotropy and a distance to Cyg X-1 of $d=2$ kpc, a value is in the middle of the range of about 1.8--2.2 kpc of \citet{zi05}, and consistent with the 1.5--2 kpc range found by \citet{cn09}. The results are shown in Fig.\ \ref{f:correlation}. (Changing $d$ would move the Cyg X-1 points diagonally.) 

We see that the Cyg X-1 dependence at 15 GHz is only slightly steeper than $p\sim 1.4$ found for H1743--322. The Cyg X-1 correlation slope for 3--9 keV is $p=1.48\pm 0.05$ (Section \ref{range}). Given the above, relatively minor, correction to the correlation slope for free-free absorption of $r\sim 0.8$, we deduce $p_{\rm intr}\sim 1.2$ for 3--9 keV. The 2.25 GHz and 8.3 GHz data are more uncertain but suggest a similar slope. Thus, Cyg X-1 appears to be intrinsically similar to H1743--322, rather than to the standard binaries showing the radio/X-ray correlation, GX 339--4 and V404 Cyg, with $p\sim 0.6$. We also note  that the normalizations of the 2.25 GHz, 8.3 GHz and 15 GHz flux distributions are very similar, which is not compatible with the 15 GHz data being strongly affected by attenuation and again suggests that the correction due to free-free absorption is relatively low. 

\subsubsection{Accretion flow models and radio emission}
\label{accretion}

\citet{coriat11} gives an extensive discussion of theoretical models that could explain $p\simeq 1.4$, and we refer the reader there for details. Assuming a scale-free jet, \citet{hs03} have obtained the following scaling between the jet power, $Q_{\rm j}$, and the flux at a given radio frequency (below the turnover $\nu$),
\begin{equation}
F_{\rm R} \propto Q_{\rm j}^\xi, \qquad \xi={2p_{\rm e} +(p_{\rm e}+6)\alpha+13\over 2(p_{\rm e}+4)}.
\label{fr_Q}
\end{equation}
Here, $p_{\rm e}$ is the index of the power-law relativistic electron distribution, $n_{\rm e}(\gamma)\propto \gamma^{-p_{\rm e}}$, and $\gamma$ is the Lorentz factor, with typical values of $p_{\rm e}\simeq 2$--3 (from observations of optically-thin parts of synchrotron spectra). For Cyg X-1  the radio photon index is $\alpha\simeq 0$, in which case $\xi\simeq 1.4$, with only a weak dependence on $p_{\rm e}$. We assume the jet power is a constant fraction, $f_{\rm j}<1$, of the accretion power, i.e., $Q_{\rm j}=f_{\rm j}\dot M c^2$. Then, $\dot M\propto F_{\rm bol}/\epsilon$, equation (\ref{mdot}). Advection models predict $\epsilon$ decreasing with the decreasing accretion rate, $\epsilon\propto \dot M^{q-1}$, where $q\sim 2$ is the index relating $F_{\rm bol}$ to $\dot M$. On the other hand, $\epsilon$ is a constant, $\sim 0.1$, in fully efficient accretion flows, and thus $q=1$. This scaling connects the radio flux to the bolometric one,
\begin{equation}
F_{\rm R}\propto F_{\rm bol}^{\xi/q}.
\label{fr_fbol}
\end{equation}
Thus, $p_{\rm intr}=\xi/q$, implying $p_{\rm intr}\simeq 1.4$ for sources with radiatively efficient accretion flows, and $p_{\rm intr}\simeq 0.7$ for sources with advective accretion flows. 

Our results of Section \ref{free-free} indicate that $p_{\rm intr}\sim 1.3$ or so through all the states of Cyg X-1 (although with with caveats due to the effect of wind absorption being not fully known), and thus its accretion flow appears to be efficient. In the hard state, this conclusion supports the same finding of \citet*{mbf09} but based on the difference in the efficiency between the average hard state and the soft state in Cyg X-1 being small. In this state, a likely model is a luminous hot accretion flow \citep{yuan01}. 

In the soft state, the hot electrons most likely form a corona above a disc, which corona is radiatively efficient, being cooled by the disc blackbody photons. Then, $q=1$ and we expect the same correlation index in the soft state for the hot electron emission ($F_{\rm hot}$) as in the hard state provided the power supplied to the corona is a constant fraction of the accretion power. Fig.\ \ref{f:hot_bol} shows that this is indeed the case, with at most a weak dependence, and the fraction equals $\sim$1/4. Both the constancy and the fractional value are consistent with theoretical expectations for the gas-pressure dominated disc \citep{merloni03,mhd03}. This is also in agreement with the finding that the disc in Cyg X-1 is in the gas-pressure dominance regime \citep{gierlinski99}. Then, the fact that the radio/X-ray correlation extends, with about the same $p$, from the hard state to the soft one (for $F_{\rm hot}$), represents then another argument for the high accretion efficiency in the hard state Cyg X-1. 

\subsubsection{X-ray jet models}
\label{xray_jet}

An alternative to the accretion flow model as an explanation of the radio/X-ray correlation is that both radio and X-rays are emitted by the jet. In particular, both could be emitted by the synchrotron emission of the same population of non-thermal electrons, with the radio and X-rays from the optically thick and thin parts, respectively. For this case, \citet{hs03} have derived,
\begin{equation}
F_{\rm R} \propto F_{\rm X}^\zeta, \qquad \zeta={2+\alpha +(3+\alpha)/(p_{\rm e}+5) \over 2+p_{\rm e}/2},
\label{fr_fx}
\end{equation}
which yields $\zeta\simeq 0.8$--0.7 for $p_{\rm e}=2$--3. Thus, this model would have required $p_{\rm intr}=\zeta\simeq 0.7$--0.8, and a large correction to $p$ due to free-free absorption, which is in conflict with our results in Section \ref{free-free}. 

Moreover, Fig.\ \ref{f:sprx} shows that the normalization of the X-ray emission in the hard state is too high by a factor of $\sim$30 for the X-rays to be due to optically thin non-thermal synchrotron emission above the turnover frequency, which is $\simeq$0.1 eV in Cyg X-1 \citep{rahoui11}. The situation becomes even worse for the intermediate state (in which the radio emission peaks), where the disagreement in the normalization at 0.1 eV is by three orders of magnitude. As discussed in Section \ref{free-free}, the effect of free-free absorption on the average 2--220 GHz spectrum can be at most a factor of $\sim$2, which cannot explain this discrepancy. This rules out non-thermal synchrotron as the origin of the X-rays. This has also been independently found by \citet{rahoui11}. We note that another ground for ruling out this model for luminous hard state of black-hole binaries is that it does not reproduce the shape of the high-energy cut-off, and it has problems reproducing the narrow distribution of the observed cut-off energies \citep{z03}. 

We note that \citet{russell10} found the X-ray spectra of the black-hole binary XTE J1550--564 to be dominated by thermal Comptonization in its luminous hard states, at $L\ga 2\times 10^{-3} L_{\rm E}$. Cyg X-1 in the hard state is relatively luminous, typically with $L\ga 0.01 L_{\rm E}$ (Z02), where $L_{\rm E}$ is the Eddington luminosity. Thus, our findings above are consistent with those of \citet{russell10}. On the other hand, the spectra of XTE J1550--564 were found to be dominated by jet non-thermal synchrotron emission at lower luminosities. A weak jet contribution, not detectable in our X-ray data, remains also possible in Cyg X-1, given the relative normalization of the radio power law and that of the X-rays, see Fig.\ \ref{f:sprx}. 

X-ray jet models for luminous states of black-hole binaries, and for Cyg X-1 in particular, have been also ruled out on several other grounds by \citet{mbf09}. In particular, they discuss the model of \citet*{mnw05}, in which the X-rays in the hard state of Cyg X-1 (and other black-hole binaries) are due to synchrotron-self-Compton by thermal electrons with $kT_{\rm e}\sim 3$--5 MeV of a very low Thomson optical depth at the jet base. \citet{mbf09} find that the parameters fitted by \citet{mnw05} strongly violate the e$^\pm$ pair equilibrium, with the self-consistent optical depth being two orders of magnitude higher than the fitted one due to the produced pairs. Furthermore, that model relies on first-order Compton scattering to fit the X-rays, and thus it requires strong fine-tuning (as noted by \citealt{y07}) to reproduce the observed spectral cut-off, which is at $\sim$100 keV in both Cyg X-1 and in most of the black-hole binaries in the hard state. Other problems with the X-ray jet models are discussed by \citet{heinz04} (effect of electron energy loss on optically-thin synchrotron spectra) and by \citet{maccarone05} (comparison of black-hole and neutron-star sources).

A number of papers have also attributed the high-energy tail observed in the soft state of X-ray binaries to optically thin non-thermal synchrotron or self-Compton emission from the jet, e.g., \citet{vadawale01}, \citet{fiocchi06}. We note that for the soft state in Cyg X-1, the extrapolation of the high-energy tail down to 1 eV is a few orders of magnitude above any possible extrapolation of the radio emission in that state. This rules out this model. In addition, it requires that the actual bolometric luminosity of the modelled sources is a few orders of magnitude above the ones determined based on the X-rays, as well as that the emission is unbeamed, which appear highly unlikely. Instead, these high-energy tails appear compatible with Compton scattering of disc blackbody photons by a hybrid electron distribution, e.g., \citet{gierlinski99}.

\section{Summary and conclusions}
\label{conclusions}

\subsection{X-rays}

We have obtained broad-band X-ray spectra of Cyg X-1 as a function of time using the broad-band monitoring by \xte/ASM together with either \gro/BATSE or \swift/BAT. Based on these spectra, we have calculated both the bolometric fluxes ($F_{\rm bol}$), and approximate values of the Comptonization flux ($F_{\rm hot}$; not including the blackbody component). We have classified the spectra into three spectral states based on their 3--12 keV photon index. We present light curves of $F_{\rm bol}$ for the available periods. The range of variability of the bolometric flux using 1-day averages is by a factor of $\simeq 10$. We find the fluxes in different states overlap, e.g., some fluxes in the hard state are higher than some in the soft state. This indicates the presence of some X-ray hysteresis in Cyg X-1, though weaker than that in low-mass X-ray binaries.  

We have studied X-ray variability patterns of Cyg X-1. In the hard state, the dominant pattern in the $\sim$10--150 keV range is of the intrinsic spectrum changing its normalization only, but with apparently more absorption at soft X-rays, causing their flux to respond to changes of the bolometric luminosity more strongly than the hard X-ray flux. In the intermediate state, there is strong spectral variability with the overall spectra changing their slope with a pivot around 20 keV. In the soft state, there is an approximate proportionality of the flux in a given energy band to $F_{\rm bol}$, but with the scatter strongly increasing with the photon energy. Still, $F_{\rm hot}/F_{\rm bol}$ was found approximately constant on average in this state, $\simeq 1/4$, see Fig.\ \ref{f:hot_bol}. 

\subsection{Radio vs.\ X-rays}

We have shown that the character of the 15 GHz radio/X-ray correlation in Cyg X-1 strongly depends on the chosen X-ray band. Results based on soft X-rays significantly differ from those based on hard X-rays. The correlation indices in the hard state vary from $p\simeq 0.9$ to $\simeq 1.8$ across the X-ray energy band. This dependence can be understood as a combination of the relationship between radio flux and bolometric flux with that between the latter and the narrow-band X-ray flux. Instead of using such narrow X-ray bands, the bolometric and Comptonization fluxes provide a much better direct measure of the underlying physics. In the hard state, the correlation index for either of them is $p\simeq 1.7$. The values of $p$ obtained based on the broad-band fluxes are larger by $\simeq$0.2--0.3 than those obtained using the commonly used X-ray band of 3--9 keV. 

We have found the radio/X-ray correlation is present in all spectral states of Cyg X-1. The radio flux peaks for relatively soft spectra, with $\Gamma\simeq 2$. For softer spectra, the radio flux drops rapidly if compared to soft X-rays, but it retains a power law dependence with hard X-rays. In particular, the radio correlation with the Comptonization flux forms a single dependence, with $p\simeq 1.7$, across all of the spectral states of Cyg X-1. This indicates that the radio jet is formed by the hot Comptonizing electrons in the accretion flow, and not by the blackbody disc. 

The radio flux in Cyg X-1 is attenuated by free-free absorption in the stellar wind from the companion, which increases the value of the observed correlation index, $p$, above the intrinsic value, $p_{\rm intr}=r p$, $r<1$. The results of \citet{zdz11} supported by our results at 2.25 GHz, 8.3 GHz and 15 GHz and their comparison with the correlation for low-mass X-ray binaries containing black holes suggest $p_{\rm intr}\sim 1.3$ or so (for the bolometric luminosity). This is similar to $p_{\rm bol}\simeq 1.4$ characteristic of efficient hot accretion flow in the framework of accretion X-ray models, implying the presence of such a flow in Cyg X-1. Based on the relative normalization of the radio and X-ray fluxes in Cyg X-1, we rule out X-ray jet models in which the X-rays are due to an extrapolation of the radio spectrum. 

\section*{ACKNOWLEDGMENTS}

We thank T. Belloni, S. Corbel, R. Narayan, M. Ostrowski, A. R. Rao and F. Yuan for valuable discussions. We also thank M. Coriat for providing us with the data used in Fig.\ \ref{f:correlation}, and the referee for valuable suggestions. This research has been supported in part by the Polish MNiSW grants N N203 581240, N N203 404939 and 362/1/N-INTEGRAL/2008/09/0. The AMI Arrays are operated by the University of Cambridge and supported by the STFC. The Green Bank Interferometer is a facility of the National Science Foundation operated by the NRAO in support of NASA High Energy Astrophysics programs. We acknowledge the use of data provided by the \xte\/ ASM  and \swift\/ BAT teams.

\appendix

\section{Radio flares from Cyg X-1}
\label{flares}

\begin{figure*}
\centerline{\includegraphics[width=8cm]{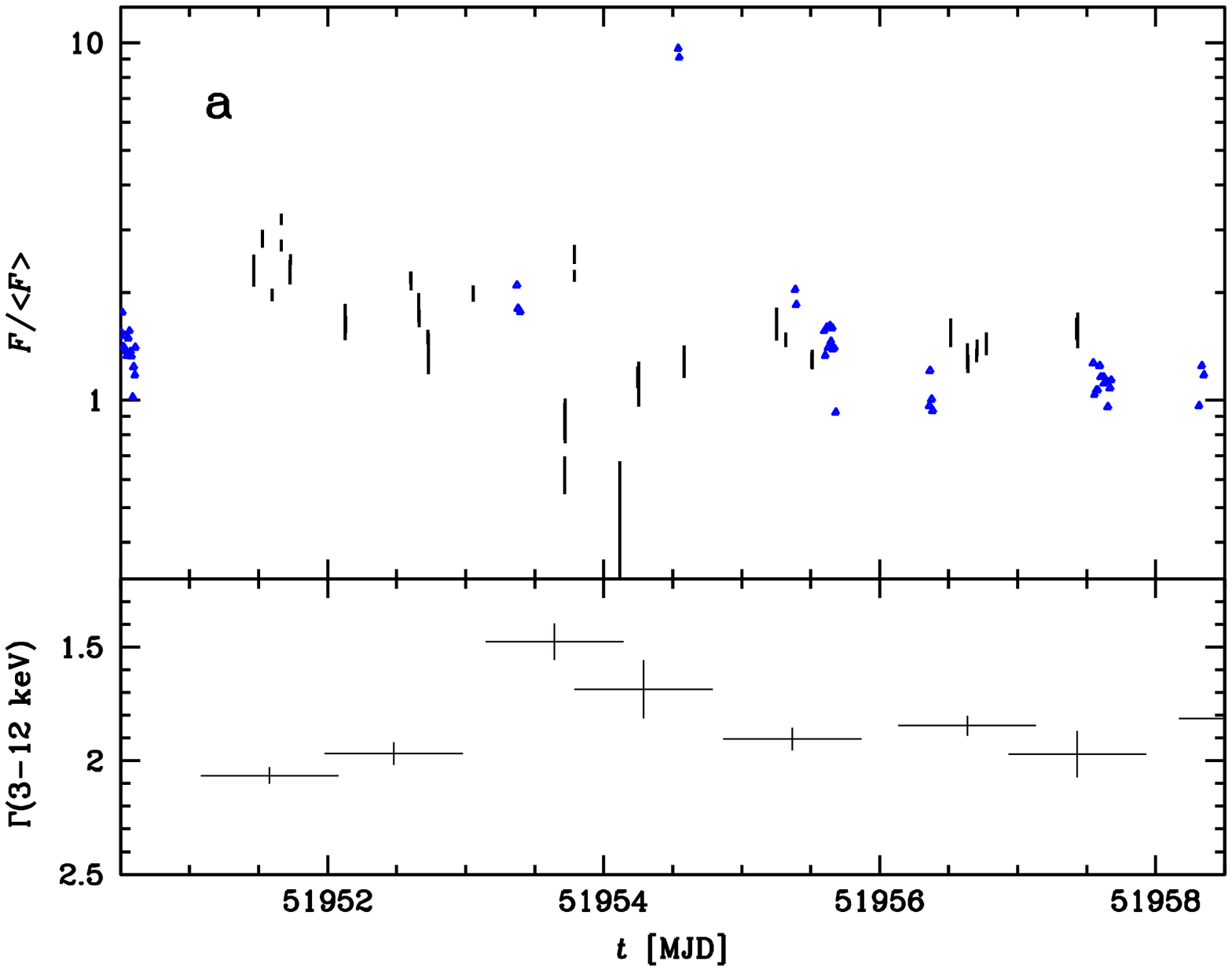}
\includegraphics[width=8cm]{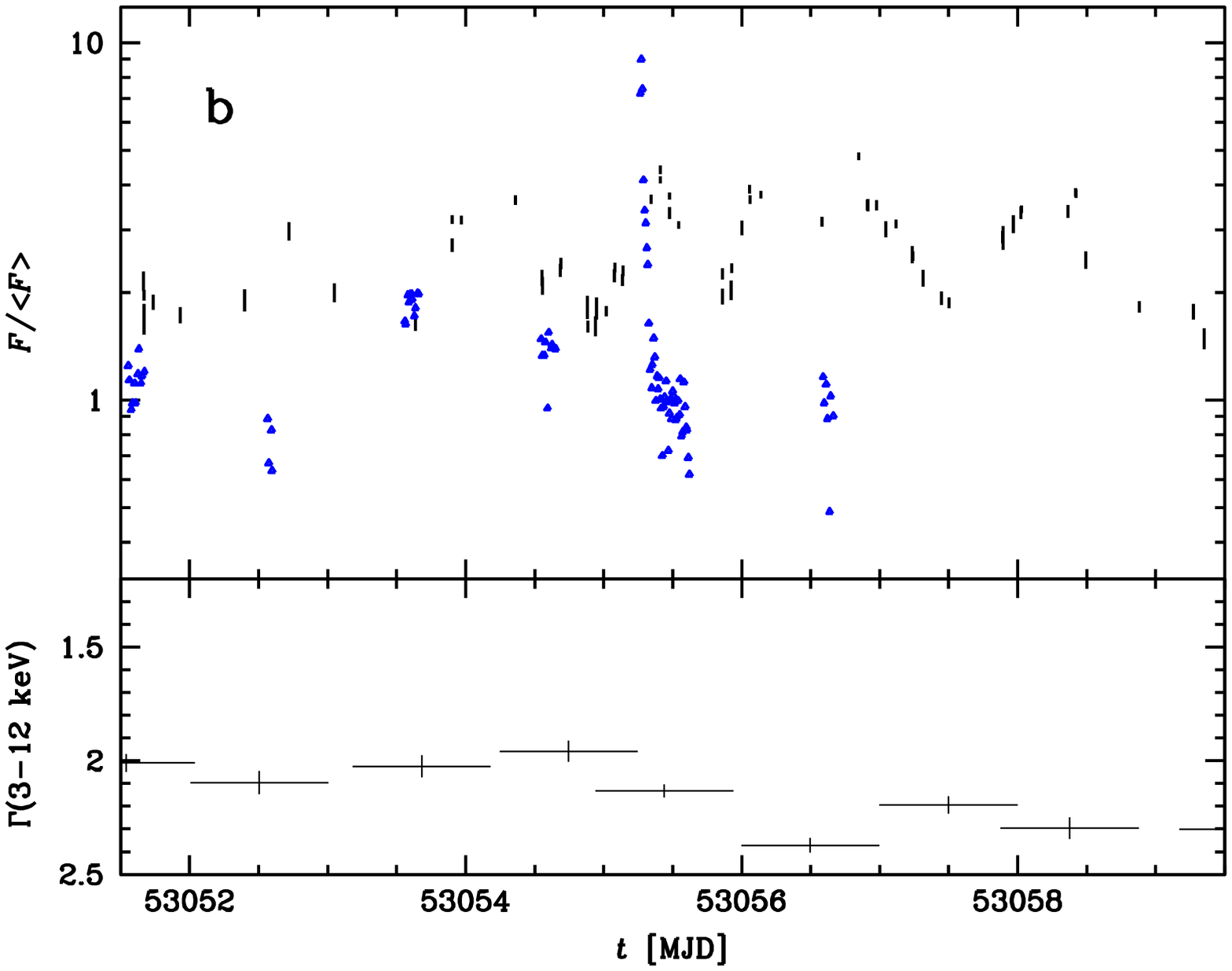}} 
\centerline{\includegraphics[width=8cm]{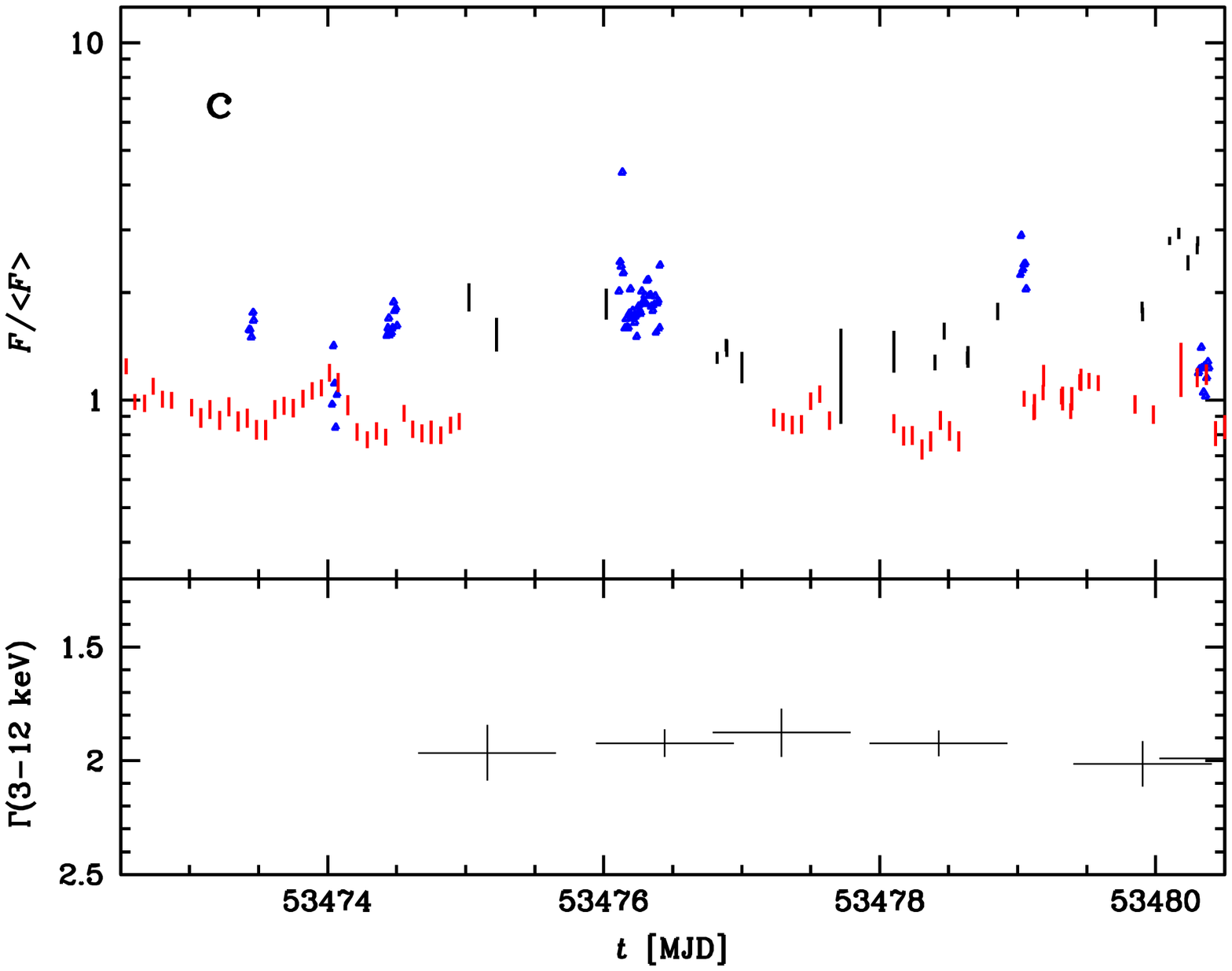}
\includegraphics[width=8cm]{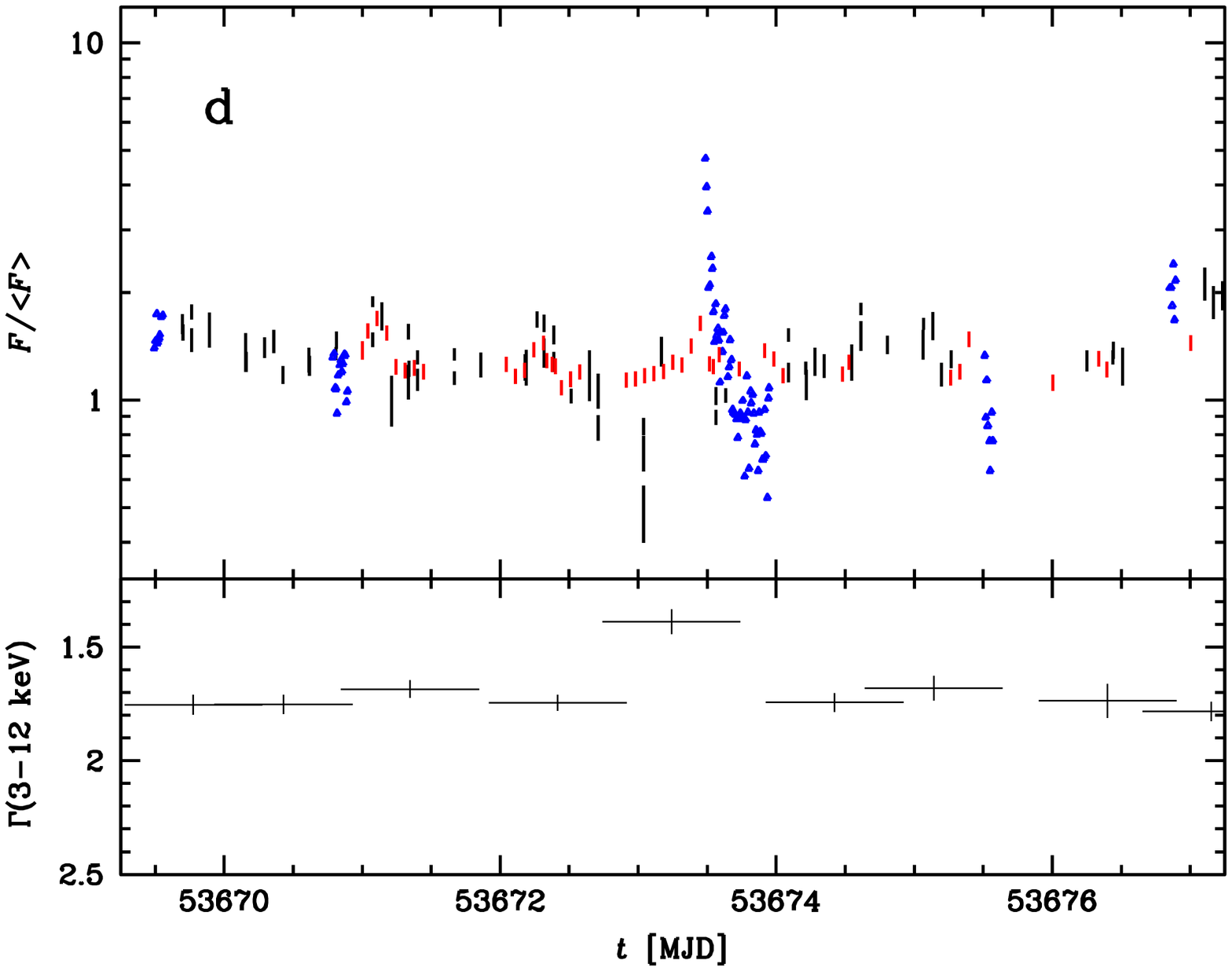}} 
\caption{The four radio flares occurring during the Ryle/AMI monitoring of Cyg X-1. The top and bottom panel for each flare gives the fluxes and the 3--12 keV photon index, respectively. The blue, black and red symbols on the top panels correspond to 15 GHz, 1.5--12 keV and 15--50 keV, respectively. All the fluxes are normalized to the corresponding average hard-state flux. 
}
\label{f:flares}
\end{figure*}

The (arithmetic) average 15-GHz flux of Cyg X-1 in its long hard state of MJD 53880--55375 is $\langle F(\rm 15\, GHz)\rangle =12.7$ mJy, and it is typically much lower during the soft states. The fractional intrinsic rms is $\sigma(\rm 15\, GHz)/\langle F(\rm 15\, GHz)\rangle=  0.35$ (ZPS11). Here, we define a flare as an event when $F_{\rm 15\, GHz}>4 \langle F_{\rm 15\, GHz}\rangle\simeq 50$ mJy, corresponding to $\simeq 8.5\sigma(\rm 15\, GHz)$. During the monitoring, see Fig.\ \ref{f:lc}, there were four such events, with the peak fluxes on MJD 51954.5396, 53055.2721 (\citealt{fender06}; note that the flare times given on the horizontal axis of their fig.\ 4, MJD 53049, and in the caption to fig.\ 4, 2005 Feb.\ 20 = MJD 53421, are incorrect), 53476.1385 \citep{wilms07} and 53673.4894. The first and fourth flares have not been reported before.

The time dependencies during 8-d intervals including the flares are shown in Fig.\ \ref{f:flares}. The BAT data shown here are from the transients web page\footnote{http://swift.gsfc.nasa.gov/docs/swift/results/transients}. Overall, we see the flares are not correlated with the X-ray emission, confirming the lack of significant short-term ($\ll $1 day) correlations in Cyg X-1 found by \citet{gleissner04}. We see that the first three flares took place during time intervals during which Cyg X-1 was in the intermediate spectral state, with $\Gamma(3$--$12\,{\rm keV})\simeq 2.0$--2.2, though the first flare itself corresponded to a transient return to the hard state, with a hardening up to $\Gamma(3$--$12\,{\rm keV})\simeq 1.5$--1.7. The fourth flare took place during an apparently normal hard state, but the flare itself was again associated with a spectral hardening, from $\Gamma(3$--$12\,{\rm keV})\simeq 1.7$--1.8 up to $\simeq 1.5$. 

\section{Calculation of the bolometric flux}
\label{bol}

\begin{figure}
\centerline{\includegraphics[width=\columnwidth]{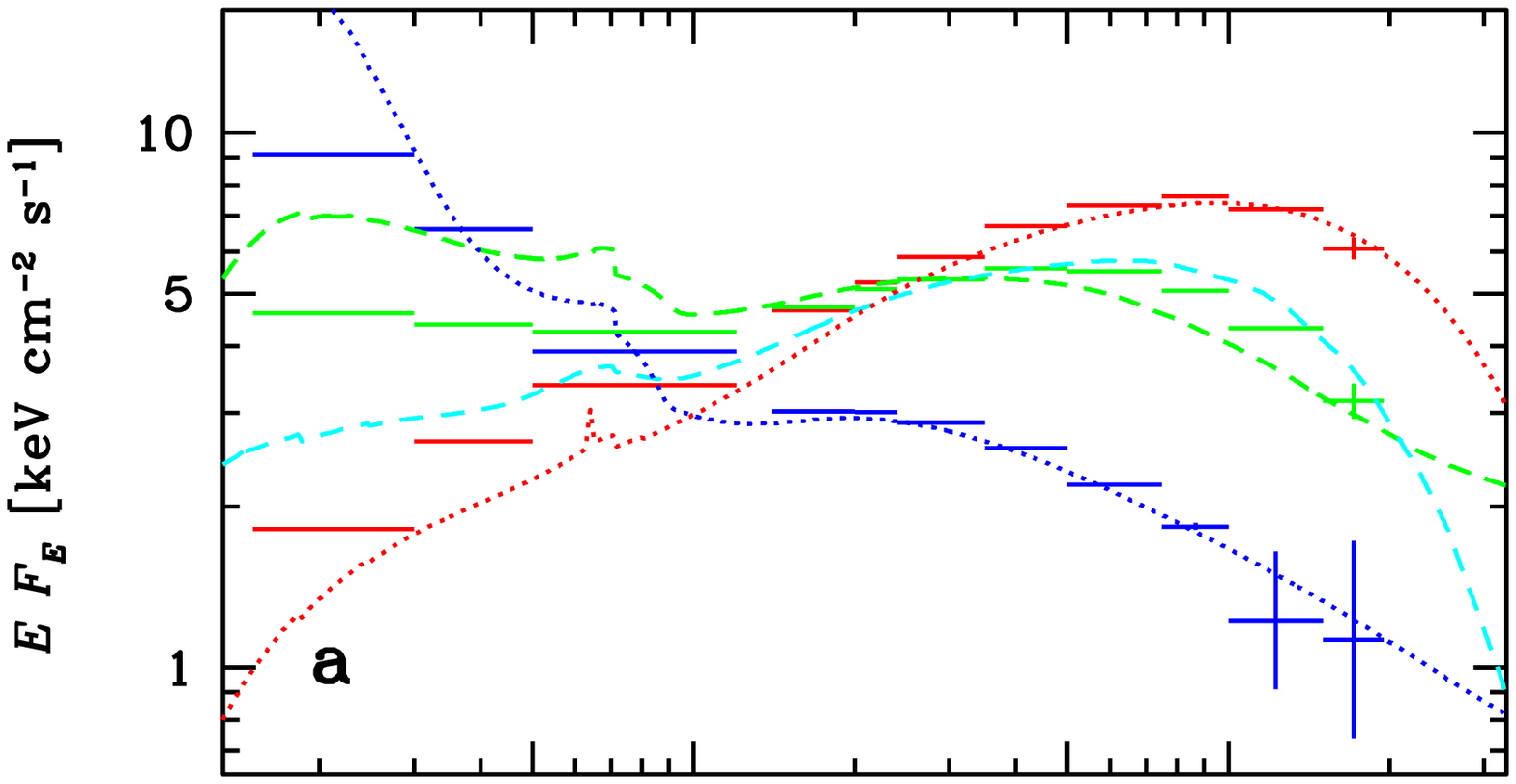}} 
\centerline{\includegraphics[width=\columnwidth]{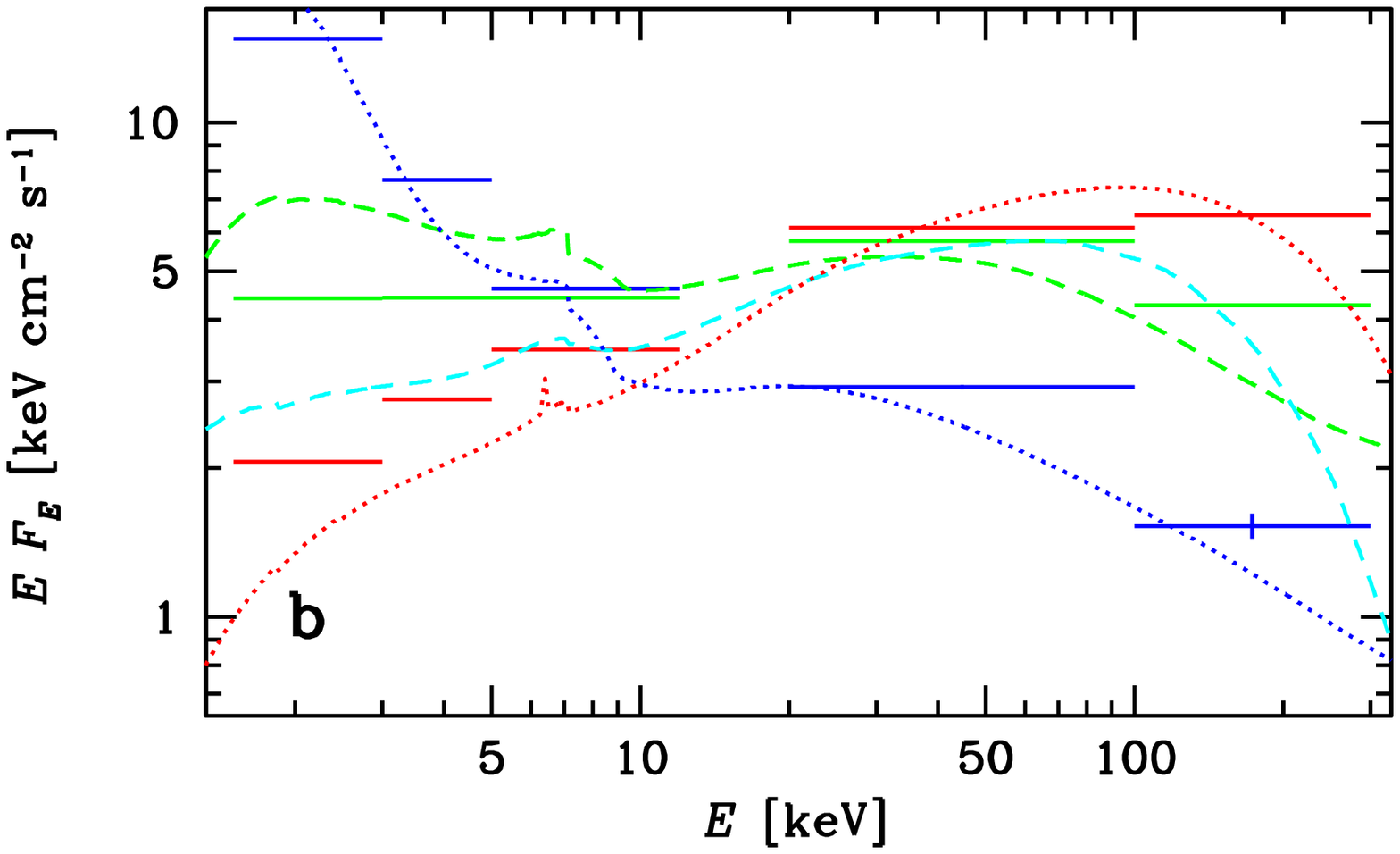}} 
\caption{
The average spectra in the hard (red error bars), intermediate (green error bars) and soft (blue error bars) states from simultaneous monitoring by (a) ASM and BAT, and (b) by ASM and BATSE, compared with best-fit models from selected pointed observations. The hard and soft state models are shown in red and blue, respectively, dotted curves, and for the intermediate state we show models for two pointed observations, in green and cyan dashed curves. See Appendix \ref{bol} for details.
}
\label{f:spectra}
\end{figure}

Here we present our method of calculating the bolometric flux based on the ASM, BATSE and BAT data. We first show a comparison of the average spectra in the three spectral spectral states, hard, intermediate and soft, defined in Section \ref{data} based on the 3--12 keV photon index, with selected pointed observations. Figs.\ \ref{f:spectra}(a--b) show the (unweighted geometric) average spectra using the simultaneous monitoring by ASM and BAT, and by ASM and BATSE, respectively. The average spectra, based on 11 and 5 channels, respectively, are then compared with model broad-band spectra fitted to pointed observations, for which we use the compilation shown in fig.\ 13 of Z02. 

For the hard state, Fig.\ \ref{f:spectra} shows the best-fit model for the \sax\/ observation on 1998 May 3--4 \citep{ds01}. We have fitted it with the thermal Comptonization model of \citet{ps96}, {\tt compps} in {\tt xspec} \citep{da01}. We find for the best fit parameters an electron temperature of $kT_{\rm e}\simeq 77$ keV, the radial Thomson optical depth in spherical geometry of $\tau_{\rm T}\simeq 2.2$, and the fractional strength of Compton reflection \citep{mz95} of 0.37, parameters typical of black-hole binaries, and of Cyg X-1 in particular.  We see a very good agreement at high energies, $\ga 20$ keV. On the other hand, that spectrum is somewhat harder than those from the ASM data. In general, we cannot expect a full agreement between average spectra selected based on $\Gamma(3$--12 keV) and any one from pointed observations of a short duration, given that the average spectra contain contributions from spectra of different shapes. In particular in the hard state, other pointed spectra are substantially softer in the ASM energy range, see, e.g., fig.\ 13 in Z02, which explains this minor discrepancy.

For the soft state, we show the \sax\/ and \gro\/ model spectrum from 1996 June, fitted with hybrid Comptonization \citep{mcconnell02}. We see relatively good agreement with our average spectra, especially with the ASM+BAT data at $E\geq 5$ keV. We see some disagreement in the 1.5--3 keV channel for the ASM+BAT spectrum, and in the normalization of the 20--300 keV channels for the ASM+BATSE spectrum. As in the hard state, this is due to the difference between observing for a short time in a given state and averaging over all data with $\Gamma(3$--12 keV) within a given range. For the intermediate state, we show models of two pointed spectra, one based on an \xte\/ spectrum of 1996 May 23 (green dashed curve), and one from a \sax\/ 1996 September 12 observation (cyan dashed curve). These two spectra correspond to the extremes of the adopted range of $\Gamma(3$--12 keV), $\simeq 2.3$, 1.9, respectively. Their average approximates relatively well the monitoring averages. 

To obtain the bolometric flux, we first add together the calculated energy fluxes in the ASM and BATSE or BAT bands. Second, we estimate the contributions from the energy intervals not covered by those instruments, and correct for absorption. We calculate the fluxes in the 12--14 keV ($<$ BAT) and 12--20 keV ($<$ BATSE) energy ranges using the observed $F(5$--12 keV) and extrapolating using the values of $\Gamma_{\rm av}$ (Section \ref{data}). We then correct the ASM fluxes for absorption by matter of cosmic composition (using the {\tt wabs} model in {\tt xspec}). We assume $N_{\rm H}=1\times 10^{22}$ cm$^{-2}$, which is the average $N_{\rm H}$ for the pointed observations of Cyg X-1 studied by \citet{i05} (using their model 0), and assume the three values of $\Gamma_{\rm av}$. We approximate the contribution from $E<1.5$ keV by assuming that the intrinsic (i.e., corrected for absorption) spectrum in the 1.5--3 keV range continues down as a power law with the values of $\Gamma_{\rm av}$ down to 0.2 keV for the hard state and to 0.4 keV for the other states, which limiting values of energy we estimated from fig.\ 1 of \citet{zg04}. The corrections for emission below 1.5 keV appear most uncertain for soft states (where the dominant part of the spectrum appears to be a disc blackbody). On the other hand the disc blackbody is weak and furthermore the contribution from $E<1.5$ keV to the bolometric flux is minor in the hard state, implying that our estimates of $F_{\rm bol}$ in this state are reasonably accurate. The contributions from energies above the range of BAT and BATSE are estimated using the shape of the \sax\/ hard-state spectrum. We find that $F(>\!\!195\,{\rm keV})\simeq 2F(150$--$195\,{\rm keV})$ (which is only $\simeq 10$ per cent of $F_{\rm bol}$) and $F(>\!\!300\,{\rm keV})\simeq 0.2F(100$--$300\,{\rm keV})$ ($\simeq 4$ per cent of $F_{\rm bol}$). These estimates are valid for the hard state, in which the spectra fall off exponentially at high energies. In soft states, the spectra continue at high energies as power laws (see, e.g., fig.\ 13 in Z02), so the correction factors would be higher. However, most of the contribution in the soft state is from soft X-rays, so that the underestimate is negligible. 

We note that in order to calculate $F_{\rm bol}$ we could have alternatively fitted the ASM and either BAT or BATSE data with some models including absorption. This would be more accurate but would have presented a number of additional problems. Whereas we can use the relatively simple thermal Comptonization in the hard state, a non-thermal electron component is required in the intermediate and soft states (e.g., \citealt{gierlinski99}), which would add a few free parameters. The fitting would have to be performed for as many as the $\sim$3000 daily spectra. Taking these issues into account, we have opted for the method based on individual-channel contributions described above. 

\section{The fitting method}
\label{fits}

We fit the X-ray/X-ray and radio/X-ray data by a power law using the method symmetric in two fitted variables, both subject to uncertainties  \citep{press92},
\begin{equation}
F_{\rm R,bol}
=F_{\rm R,bol,0} \left(F_{\rm X}\over \langle F_{\rm X}\rangle\right)^p,\qquad F_{\rm R}>F_{\rm R,min}.
\label{fit}
\end{equation}
Here $F_{\rm X}$ is either the flux in one of the X-ray energy intervals or a broad-band X-ray flux, $F_{\rm R,bol}$ is either a radio flux or the X-ray bolometric flux, and $\langle F_{\rm X}\rangle$ is the geometric average of the X-ray flux, which is chosen as the normalization in order to minimize the fit uncertainties. The parameters obtained are $F_{\rm R,bol,0}$, which is either the radio or bolometric X-ray flux at $\langle F_{\rm X}\rangle$, and the power-law index, $p$. The logarithms of $F_{\rm R,bol}$ and $F_{\rm X}$ are fitted. In the case of radio/X-ray correlations, we fit the data only with $F_{\rm R}>F_{\rm R,min}$, below which the correlations found here break down, probably due to large errors on the radio measurements. The fit parameters are given in Table \ref{t:fit}.

We encounter here an important issue of the uncertainties on the fitted parameters, $F_{\rm R,bol,0}$ and $p$. The probability calculated by this method assumes that the spread of the points is solely due to measurement errors whereas it is, as usually the case in astrophysics, due to both the intrinsic variability and the measurement errors. If only the errors are included, the fit probabilities are very close to null, which renders invalid the estimates on the fit uncertainties. Therefore, we assume fractional uncertainties, $\delta_{\rm x}$, $\delta_{\rm y}$, on all the measurements such that the fit probability becomes $\simeq$0.5, or equivalently, $\chi^2_\nu\simeq 1$. This takes into account the actual spread of the data points due to intrinsic variability. The ratio between the errors in the $x$ and $y$ directions is taken to be the ratio  of the observed spread of the points, $\delta_{\rm y}/\delta_{\rm x}=\sigma_{\rm y}/\sigma_{\rm x}$, where $\sigma_{\rm x,y}$ is the standard deviation (in the log space) of the $x$ or $y$ points, respectively. Since the measurement errors are typically much smaller than the spread of the points we neglect them and include only the errors as above. (This method is similar to that used in \citealt{mhd03} except that they assumed $\delta_{\rm x}=\delta_{\rm y}$.) Under our assumptions, the best fit index corresponds to the slope of the covariance ellipse, $p\simeq \sigma_{\rm y}/\sigma_{\rm x}$. The best-fit parameter values depend only weakly on $\delta$ as long as their ratio is fixed as above. The fractional errors required for $\chi^2_\nu\simeq 1$ are in general $<0.3$ except for fits involving the 150--195 keV energy band, where they are required to be slightly larger. 

For tight correlations, such as between the X-ray flux in a given energy band, the fitted values of $p$ weakly depend on the $\delta_{\rm x}/\delta_{\rm y}$ ratio, and they are almost the same if, e.g., $\delta_{\rm x}=\delta_{\rm y}$ were assumed. On the other hand, if the spread is large, as in the radio/X-ray correlations, the fitted values depend sensitively on the adopted $\delta_{\rm x}/\delta_{\rm y}$. However, our method above does reproduce the actual slope of the dependence, under the assumption that it is linear in the log space and that all points have equal relative weight. 

\label{lastpage}


\begin{thebibliography}{}

\bibitem[\protect\citeauthoryear{Barthelmy et al.}{2005}]{barthelmy05}
Barthelmy S. D. et al., 2005, Space Sci. Rev., 120, 143

\bibitem[\protect\citeauthoryear{Blandford \& Konigl}{1979}]{bk79} 
Blandford R.~D., Konigl A., 1979, ApJ, 232, 34 

\bibitem[\protect\citeauthoryear{Bradt, Rothschild \& Swank}{Bradt et al.}{1993}]{brs93}
Bradt H. V., Rothschild R. E., Swank J. H., 1993, A\&AS, 97, 355

\bibitem[\protect\citeauthoryear{Churazov, Gilfanov \& Revnivtsev}{Churazov et al.}{2001}]{cgr01} 
Churazov E., Gilfanov M., Revnivtsev M., 2001, MNRAS, 321, 759 

\bibitem[\protect\citeauthoryear{Caballero-Nieves et al.}{2009}]{cn09} 
Caballero-Nieves S.~M., et al., 2009, ApJ, 701, 1895 

\bibitem[\protect\citeauthoryear{Corbel et al.}{2000}]{corbel00} 
Corbel S., Fender R.~P., Tzioumis A.~K., Nowak M., McIntyre V., Durouchoux P., Sood R., 2000, A\&A, 359, 251 

\bibitem[\protect\citeauthoryear{Corbel et al.}{2003}]{corbel03} 
Corbel S., Nowak M.~A., Fender R.~P., Tzioumis A.~K., Markoff S., 2003, A\&A, 400, 1007 

\bibitem[\protect\citeauthoryear{Corbel et al.}{2004}]{corbel04} 
Corbel S., Fender R.~P., Tomsick J.~A., Tzioumis A.~K., Tingay S., 2004, 
ApJ, 617, 1272 

\bibitem[\protect\citeauthoryear{Corbel, Koerding \& Kaaret}{Corbel et al.}{2008}]{corbel08} 
Corbel S., Koerding E., Kaaret P., 2008, MNRAS, 389, 1697 

\bibitem[\protect\citeauthoryear{Coriat et al.}{2009}]{coriat09} 
Coriat M., Corbel S., Buxton M.~M., Bailyn C.~D., Tomsick J.~A., K{\"o}rding E., Kalemci E., 2009, MNRAS, 400, 123 

\bibitem[\protect\citeauthoryear{Coriat et al.}{2011}]{coriat11} 
Coriat M., et al., 2011, MNRAS, in press, arXiv:1101.5159 

\bibitem[\protect\citeauthoryear{Di Salvo et al.}{2001}]{ds01} 
Di Salvo T., Done C., \.Zycki P. T., Burderi L., Robba N. R., 2001, ApJ, 547, 1024

\bibitem[\protect\citeauthoryear{Done, Gierli{\'n}ski \& Kubota}{Done et al.}{2007}]{dgk07} 
Done C., Gierli{\'n}ski M., Kubota A., 2007, A\&ARv, 15, 1 

\bibitem[\protect\citeauthoryear{Dorman \& Arnaud}{2001}]{da01}
Dorman B., Arnaud K. A., 2001, in Harnden Jr. F. R., Primini F. A., Payne H. E., eds, ASP Conf.\ Ser.\ Vol.\ 238, Astronomical Data Analysis Software and Systems X. Astron.\ Soc.\ Pac., San Francisco, p.\ 415

\bibitem[\protect\citeauthoryear{Dunn et al.}{2008}]{dunn08} 
Dunn R.~J.~H., Fender R.~P., K{\"o}rding E.~G., Cabanac C., Belloni T., 
2008, MNRAS, 387, 545 

\bibitem[\protect\citeauthoryear{Falcke, K{\"o}rding \& Markoff}{Falcke et al.}{2004}]{falcke04} 
Falcke H., K{\"o}rding E., Markoff S., 2004, A\&A, 414, 895 

\bibitem[\protect\citeauthoryear{Fender et al.}{2000}]{fender00}
Fender, R. P., Pooley, G. G., Durouchoux, P., Tilanus, R. P. J., Brocksopp, C., 2000, MNRAS, 312, 853

\bibitem[\protect\citeauthoryear{Fender, Belloni \& Gallo}{Fender et al.}{2004}]{fbg04} 
Fender R.~P., Belloni T.~M., Gallo E., 2004, MNRAS, 355, 1105 

\bibitem[\protect\citeauthoryear{Fender et al.}{2006}]{fender06} 
Fender R.~P., Stirling A.~M., Spencer R.~E., Brown I., Pooley G.~G., Muxlow T.~W.~B., Miller-Jones J.~C.~A., 2006, MNRAS, 369, 603 

\bibitem[\protect\citeauthoryear{Fiocchi et al.}{2006}]{fiocchi06} 
Fiocchi M., Bazzano A., Ubertini P., Jean 
P., 2006, ApJ, 651, 416 

\bibitem[\protect\citeauthoryear{Gallo, Fender \& Pooley}{Gallo et al.}{2003}]{gfp03}
Gallo E., Fender R.~P., Pooley G.~G., 2003, MNRAS, 344, 60

\bibitem[\protect\citeauthoryear{Gallo et al.}{2004}]{gallo04} 
Gallo E., Corbel S., Fender R.~P., Maccarone T.~J., Tzioumis A.~K., 2004, 
MNRAS, 347, L52 

\bibitem[\protect\citeauthoryear{Georganopoulos, Aharonian \& Kirk}{Georganopoulos et al.}{2002}]{georganopoulos02}
Georganopoulos M., Aharonian F.~A., Kirk J.~G., 2002, A\&A, 388, L25

\bibitem[\protect\citeauthoryear{Gierli{\'n}ski \& Done}{2004}]{gd04} 
Gierli{\'n}ski M., Done C., 2004, MNRAS, 347, 885 

\bibitem[\protect\citeauthoryear{Gierli{\'n}ski, Zdziarski \& Done}{Gierli{\'n}ski et al.}{2011}]{gzd11} 
Gierli{\'n}ski M., Zdziarski A.~A., Done C., 2011, in Frontier Objects in Astrophysics and Particle Physics 2010, Italian Phys.\ Soc., in press (arXiv:1011.5840) 

\bibitem[\protect\citeauthoryear{Gierli\'nski et al.}{1997}]{gierlinski97}
Gierli\'nski M., Zdziarski A. A., Done C., Johnson W. N., Ebisawa K., Ueda Y., Haardt F., Phlips B. F., 1997, MNRAS, 288, 958

\bibitem[\protect\citeauthoryear{Gierli{\'n}ski et al.}{1999}]{gierlinski99} 
Gierli{\'n}ski M., Zdziarski A.~A., Poutanen J., Coppi P.~S., Ebisawa K., Johnson W.~N., 1999, MNRAS, 309, 496 

\bibitem[\protect\citeauthoryear{Gies et al.}{2003}]{gies03} 
Gies D. R. et al., 2003, ApJ, 583, 424

\bibitem[\protect\citeauthoryear{Gies et al.}{2008}]{gies08} 
Gies D.~R., et al., 2008, ApJ, 678, 1237 

\bibitem[\protect\citeauthoryear{Gleissner et al.}{2004}]{gleissner04} 
Gleissner T., et al., 2004, A\&A, 425, 1061 

\bibitem[\protect\citeauthoryear{Harmon et al.}{2002}]{harmon02} 
Harmon B. A. et al., 2002, ApJS, 138, 149

\bibitem[\protect\citeauthoryear{Heinz}{2004}]{heinz04} 
Heinz S., 2004, MNRAS, 355, 835 

\bibitem[\protect\citeauthoryear{Heinz \& Sunyaev}{2003}]{hs03} 
Heinz S., Sunyaev R.~A., 2003, MNRAS, 343, L59 

\bibitem[\protect\citeauthoryear{Hjalmarsdotter et al.}{2009}]{hj09} 
Hjalmarsdotter L., Zdziarski A.~A., Szostek A., Hannikainen D.~C., 2009, MNRAS, 392, 251

\bibitem[\protect\citeauthoryear{Ibragimov et al.}{2005}]{i05} 
Ibragimov A., Poutanen J., Gilfanov M., Zdziarski A.~A., Shrader C.~R., 2005, MNRAS, 362, 1435

\bibitem[\protect\citeauthoryear{Jourdain \& Roques}{2009}]{jr09} 
Jourdain E., Roques J.~P., 2009, ApJ, 704, 17 

\bibitem[\protect\citeauthoryear{Lachowicz et al.}{2006}]{l06}
Lachowicz P., Zdziarski A. A., Schwarzenberg-Czerny A., Pooley G. G., 
Kitamoto S., 2006, MNRAS, 368, 1025

\bibitem[\protect\citeauthoryear{Levine et al.}{1996}]{levine96}
Levine A. M., Bradt H., Cui W., Jernigan J. G., Morgan E. H.,
Remillard R., Shirey R. E., Smith D. A., 1996, ApJ, 469, L33

\bibitem[\protect\citeauthoryear{Maccarone}{2005}]{maccarone05} 
Maccarone T.~J., 2005, MNRAS, 360, L68 

\bibitem[\protect\citeauthoryear{Magdziarz \& Zdziarski}{1995}]{mz95}
Magdziarz P., Zdziarski A. A., 1995, MNRAS, 273, 837

\bibitem[\protect\citeauthoryear{Malzac, Belmont \& Fabian}{Malzac et al.}{2009}]{mbf09} 
Malzac J., Belmont R., Fabian A.~C., 2009, MNRAS, 400, 1512 

\bibitem[\protect\citeauthoryear{Markoff et al.}{2003}]{markoff03}
Markoff S., Nowak M., Corbel S., Falcke H., Fender R., 2003, A\&A, 397, 645

\bibitem[\protect\citeauthoryear{Markoff, Nowak \& Wilms}{Markoff et al.}{2005}]{mnw05} 
Markoff S., Nowak M.~A., Wilms J., 2005, ApJ, 635, 1203 

\bibitem[\protect\citeauthoryear{Markwardt et al.}{2005}]{m05} 
Markwardt C.~B., Tueller J., Skinner G.~K., Gehrels N., Barthelmy S.~D., Mushotzky R.~F., 2005, ApJ, 633, L77 

\bibitem[\protect\citeauthoryear{McConnell et al.}{2002}]{mcconnell02} 
McConnell M.~L., et al., 2002, ApJ, 572, 984

\bibitem[\protect\citeauthoryear{Merloni}{2003}]{merloni03} 
Merloni A., 2003, MNRAS, 341, 1051 

\bibitem[\protect\citeauthoryear{Merloni \& Fabian}{2002}]{mf02} 
Merloni A., Fabian A.~C., 2002, MNRAS, 332, 165 

\bibitem[\protect\citeauthoryear{Merloni, Heinz \& di Matteo}{Merloni et al.}{2003}]{mhd03} 
Merloni A., Heinz S., di Matteo T., 2003, MNRAS, 345, 1057 

\bibitem[\protect\citeauthoryear{Mirabel et al.}{1996}]{mirabel96} 
Mirabel I.~F., Claret A., Cesarsky C.~J., Boulade O., Cesarsky D.~A., 1996, A\&A, 315, L113 

\bibitem[\protect\citeauthoryear{Mirabel et al.}{1998}]{mirabel98}
Mirabel I.~F., Dhawan V., Chaty S., Rodriguez L.~F., Marti J., Robinson C.~R.,
Swank J., Geballe T., 1998, A\&A, 330, L9

\bibitem[\protect\citeauthoryear{Persi et al.}{1980}]{persi80} 
Persi P., Ferrari-Toniolo M., Grasdalen G.~L., Spada G., 1980, A\&A, 92, 238 

\bibitem[\protect\citeauthoryear{Pooley, Fender \& Brocksopp}{Pooley et al.}{1999}]{pfb99} 
Pooley G.~G., Fender R.~P., Brocksopp C., 1999, MNRAS, 302, L1 

\bibitem[\protect\citeauthoryear{Poutanen \& Svensson}{1996}]{ps96}
Poutanen J., Svensson R., 1996, ApJ, 470, 249

\bibitem[\protect\citeauthoryear{Poutanen, Zdziarski \& Ibragimov}{Poutanen et al.}{2008}]{pzi08} 
Poutanen J., Zdziarski A.~A., Ibragimov A., 2008, MNRAS, 389, 1427

\bibitem[\protect\citeauthoryear{Press et al.}{1992}]{press92}
Press W. H., Teukolsky S. A., Vetterling W. T., Flannery B.
P., 1992, Numerical Recipes. Cambridge Univ.\ Press, Cambridge

\bibitem[\protect\citeauthoryear{Rahoui et al.}{2011}]{rahoui11} 
Rahoui F., Lee J.~C., Heinz S., Hines D.~C., Pottschmidt K., Wilms J., 
Grinberg V., 2011, ApJ, in press, arXiv:1105.0336 

\bibitem[\protect\citeauthoryear{Rao \& Vadawale}{2001}]{rv01} 
Rao A.~R., Vadawale S.~V., 2001, ESA SP 459, 517 

\bibitem[\protect\citeauthoryear{Rushton et al.}{2011}]{rushton11} 
Rushton A., et al., 2011, in Proceedings of Science, in press, arXiv:1101.3322 

\bibitem[\protect\citeauthoryear{Russell et al.}{2010}]{russell10} 
Russell D.~M., Maitra D., Dunn R.~J.~H., Markoff S., 2010, MNRAS, 405, 1759 

\bibitem[\protect\citeauthoryear{Soleri et al.}{2010}]{soleri10} 
Soleri P., et al., 2010, MNRAS, 406, 1471 

\bibitem[\protect\citeauthoryear{Stirling et al.}{2001}]{stirling01} 
Stirling A.~M., Spencer R.~E., de la Force C.~J., Garrett M.~A., Fender R.~P., Ogley R.~N., 2001, MNRAS, 327, 1273 

\bibitem[\protect\citeauthoryear{Szostek \& Zdziarski}{2007}]{sz07} 
Szostek A., Zdziarski A.~A., 2007, MNRAS, 375, 793 

\bibitem[\protect\citeauthoryear{Szostek \& Zdziarski}{2008}]{sz08} 
Szostek A., Zdziarski A.~A., 2008, MNRAS, 386, 593 

\bibitem[\protect\citeauthoryear{Szostek, Zdziarski \& McCollough}{Szostek et al.}{2008}]{szm08} 
Szostek A., Zdziarski A.~A., McCollough M.~L., 2008, MNRAS, 388, 1001 

\bibitem[\protect\citeauthoryear{Vadawale, Rao \& Chakrabarti}{Vadawale et al.}{2001}]{vadawale01}
Vadawale S.~V., Rao A.~R., Chakrabarti S.~K., 2001, A\&A, 372, 793

\bibitem[\protect\citeauthoryear{Wilms et al.}{2007}]{wilms07} 
Wilms J., Pottschmidt K., Pooley G.~G., Markoff S., Nowak M.~A., Kreykenbohm I., Rothschild R.~E., 2007, ApJ, 663, L97 

\bibitem[\protect\citeauthoryear{Yuan}{2001}]{yuan01} 
Yuan F., 2001, MNRAS, 324, 119 

\bibitem[\protect\citeauthoryear{Yuan \& Cui}{2005}]{yc05} 
Yuan F., Cui W., 2005, ApJ, 629, 408 

\bibitem[\protect\citeauthoryear{Yuan, Cui, \& Narayan}{Yuan et al.}{2005}]{ycn05} 
Yuan F., Cui W., Narayan R., 2005, ApJ, 620, 905 

\bibitem[\protect\citeauthoryear{Yuan et al.}{2007}]{y07}
Yuan F., Zdziarski A.~A., Xue Y., Wu X.-B., 2007, ApJ, 659, 541

\bibitem[\protect\citeauthoryear{Yuan et al.}{2009}]{y09} 
Yuan F., Lin J., Wu K., Ho L.~C., 2009, MNRAS, 395, 2183 

\bibitem[\protect\citeauthoryear{Zdziarski}{2011}]{zdz11}
Zdziarski A. A., 2011, MNRAS, submitted, arXiv:1105.4291

\bibitem[\protect\citeauthoryear{Zdziarski \& Gierli\'nski}{2004}]{zg04}
Zdziarski A. A., Gierli\'nski M., 2004, Progr.\ Theor.\ Phys.\ Suppl., 155, 99

\bibitem[\protect\citeauthoryear{Zdziarski et al.}{2002}]{z02}
Zdziarski A. A., Poutanen J., Paciesas W. S., Wen L., 2002, ApJ, 578, 357 (Z02)

\bibitem[\protect\citeauthoryear{Zdziarski et al.}{2003}]{z03} 
Zdziarski A.~A., Lubi{\'n}ski P., Gilfanov M., Revnivtsev M., 2003, MNRAS, 342, 355 

\bibitem[\protect\citeauthoryear{Zdziarski et al.}{2004}]{z04}
Zdziarski A. A., Gierli\'nski M., Miko{\l}ajewska J., Wardzi\'nski G., Smith D. M., Harmon B. A., Kitamoto S., 2004, MNRAS, 351, 791

\bibitem[\protect\citeauthoryear{Zdziarski, Pooley \& Skinner}{Zdziarski et al.}{2011}]{z11} 
Zdziarski A.~A., Pooley G.~G., Skinner G.~K., 2011, MNRAS, 412, 1985 (ZPS11)

\bibitem[\protect\citeauthoryear{Zi\'o{\l}kowski }{2005}]{zi05}
Zi\'o{\l}kowski J., 2005, MNRAS, 358, 851

\end{thebibliography}
\end{document}